\documentclass[twocolumn,twocolappendix]{aastex63}
\usepackage{newtxtext,newtxmath}
\usepackage[T1]{fontenc}
\usepackage{ae,aecompl}
\usepackage{comment}
\usepackage{graphicx,tabularx}	
\usepackage{amsmath}	
\usepackage{amssymb}	
\usepackage{booktabs}
\usepackage[inline,shortlabels]{enumitem}
\setlist{itemjoin* = { and\enspace}}
\usepackage{color}
\usepackage{multirow}
\usepackage[figuresright]{rotating}
\usepackage{fontawesome5}
\defcitealias{2021arXiv210503344Z}{Paper I} 
\defcitealias{zhao21cmdelfi}{Z22b}

\shorttitle{Reionization parameter inference}
\shortauthors{Zhao et al.}

\graphicspath{{./}{figures/}}

\begin{document}

\title{Simulation-based Inference of Reionization Parameters from 3D Tomographic 21 cm Light-cone Images - II: Application of Solid Harmonic Wavelet Scattering Transform}

\correspondingauthor{Xiaosheng Zhao, Yi Mao}\email{zhaoxs18@mails.tsinghua.edu.cn (XZ), ymao@tsinghua.edu.cn (YM)}

\author[0000-0002-8328-1447]{Xiaosheng Zhao}
\affiliation{Department of Astronomy, Tsinghua University, Beijing 100084, China}
\affiliation{Sorbonne Universit\'e, CNRS, UMR 7095, Institut d'Astrophysique de Paris (IAP), 98 bis bd Arago, 75014 Paris, France}

\author[0000-0002-1301-3893]{Yi Mao}
\affiliation{Department of Astronomy, Tsinghua University, Beijing 100084, China}

\author[0000-0003-3858-6361]{Shifan Zuo}
\affiliation{National Astronomical Observatories, Chinese Academy of Sciences, Beijing 100101, China}
\affiliation{Key Laboratory of Radio Astronomy and Technology, Chinese Academy of Sciences, A20 Datun Road, Chaoyang District, Beijing, 100101, P. R. China}

\author[0000-0002-5854-8269]{Benjamin D. Wandelt}
\affiliation{Sorbonne Universit\'e, CNRS, UMR 7095, Institut d'Astrophysique de Paris (IAP), 98 bis bd Arago, 75014 Paris, France}
\affiliation{Sorbonne Universit\'e, Institut Lagrange de Paris (ILP), 98 bis bd Arago, 75014 Paris, France}
\affiliation{Center for Computational Astrophysics, Flatiron Institute, 162 5th Avenue, New York, NY 10010, USA} 

\begin{abstract}

The information regarding how the intergalactic medium is reionized by astrophysical sources is contained in the tomographic three-dimensional 21~cm images from the epoch of reionization. In \citet{2021arXiv210503344Z} (``Paper I''), we demonstrated for the first time that density estimation likelihood-free inference (DELFI) can be applied efficiently to perform a Bayesian inference of the reionization parameters from the 21~cm images. Nevertheless, the 3D image data needs to be compressed into informative summaries as the input of DELFI by, e.g., a trained 3D convolutional neural network (CNN) as in Paper I ({\tt DELFI-3D CNN}). Here in this paper, we introduce an alternative data compressor, the solid harmonic wavelet scattering transform (WST), which has a similar, yet fixed (i.e.\ no training), architecture to CNN, but we show that this approach (i.e.\ solid harmonic WST with DELFI) outperforms earlier analyses based on 3D 21~cm images using {\tt DELFI-3D CNN} in terms of credible regions of parameters. Realistic effects, including thermal noise and residual foreground after removal, are also applied to the mock observations from the Square Kilometre Array (SKA). We show that under the same inference strategy using DELFI, the 21~cm image analysis with solid harmonic WST outperforms the 21~cm power spectrum analysis. This research serves as a proof of concept, demonstrating the potential to harness the strengths of WST and simulation-based inference to derive insights from future 21~cm light-cone image data.

\end{abstract}

\keywords{Reionization (1383), H I line emission (690), Astrostatistics (1882), Bayesian statistics (1900), Wavelet analysis (1918)}

\section{Introduction} 
\label{sec:intro}
The intensity mapping of the 21~cm line associated with the spin-flip transition of H~I atoms is a promising probe of the epoch of reionization (EoR; \citealt{Furlanetto2006}). It contains information regarding when and how the intergalactic medium (IGM) was heated and reionized by the first luminous objects. Upper limits of the 21~cm power spectrum from the EoR \citep{2013MNRAS.433..639P, Parsons_2014, 2015ApJ...801...51J, 2020MNRAS.493.1662M, 2020MNRAS.493.4711T,2021MNRAS.505.4775Y, 2022ApJ...925..221A} have been placed by ongoing interferometric array experiments, including the Precision Array for Probing the Epoch of Reionization (PAPER; \citealp{Parsons2010}), the Murchison Widefield Array (MWA; \citealp{Tingay2013}), the LOw Frequency ARray (LOFAR; \citealp{Haarlem2013}), and the Giant Metrewave Radio Telescope (GMRT; \citealp{2017A&A...598A..78I}). In the foreseeable future, the first measurements of the 21~cm power spectrum from the EoR will be very likely achieved by upcoming array experiments including the Hydrogen Epoch of Reionization Array (HERA; \citealp{DeBoer2017}) and the Square Kilometre Array (SKA; \citealp{Mellema2013}). Furthermore, the SKA will have the exciting promise of mapping the three-dimensional (3D) tomographic light-cone images of the 21~cm brightness temperature from the EoR with high sensitivity. 

The 21~cm signal is non-Gaussian due to reionization patchiness. Therefore, the 3D light-cone images of the 21~cm signal contain more information than the power spectrum statistic, which is unlike the traditionally well-studied case of Gaussian signals in the cosmic microwave background (CMB). As such, it is of key importance for 21~cm observers to understand how to optimally extract the information in the 3D 21~cm images. The 3D image data needs to be compressed into informative summaries inevitably because it is technically very challenging to process the high-dimensional image data directly. For this purpose, several works have proposed to apply the convolutional neural networks (CNNs) to compress the 2D 21~cm image slices \citep{Gillet2019} or 3D 21~cm light-cone images (\citealt{2021arXiv210503344Z}, hereafter referred to as ``\citetalias{2021arXiv210503344Z}''; \citealt{2022MNRAS.509.3852P}; \citealt{2022MNRAS.511.3446N}) into data summaries. However, the practical applications of CNN-based methods are generically computationally expensive both in generating a large volume of simulation data that are needed for training the networks and in the process of training and optimization itself. Even with so much ``engineering'', the fine-tuned networks may still be sub-optimal (see, e.g.\ \citetalias{2021arXiv210503344Z}; \citealt{2022MNRAS.509.3852P}). 

To mitigate these problems, it has been proposed to inject the inductive bias into CNNs with scattering transform \citep{mallat2012group,allys2019rwst,cheng2020new,2022mla..confE..40P} and utilize the scattering transform to construct scattering or wavelet networks \citep{2021arXiv210709539G, 2022mla..confE..40P}. The scattering transform employs the filters that have well-behaved mathematical structures, e.g.\ the Morlet filters \citep{mallat2012group,10.1093/mnras/stw1310}, and under its unique definition, exploits the modulus nonlinearities and hierarchical structures --- similar to the multi-layers in the CNN --- which allows it to extract information across multi-scales. Compared with CNNs, the scattering transform has fixed filter parameters, and therefore do {\it not} need to be trained, which is a significant advantage against the CNNs. 
Recently, the scattering transform has been extended to 3D applications. For example, the harmonic-related wavelets are introduced to infer the molecular properties \citep{eickenberg2017solid,eickenberg2018solid}, cosmological parameters \citep{2021ApJ...910..122S, 2022PhRvD.105j3534V, 2022MNRAS.517.1625C, 2022arXiv220407646E}, and the CMB B-mode \citep{2022MNRAS.510L...1J}. Specifically, \citet{2022arXiv220407646E} employs the first-order wavelet-based features and shows the advantage of harmonic wavelets against the isotropic and oriented ones. In the context of extracting the information from the 3D 21~cm light-cone (i.e.\ ``light-cuboids''), in this paper, we apply the solid harmonic wavelet scattering transform (WST; \citealp{eickenberg2017solid,eickenberg2018solid}) to compress the 3D image data. The solid harmonic WST injects the inductive bias into 3D CNNs with both 3D solid harmonic wavelets and the scattering transform which outputs multiple-order wavelet-based features. 

Traditionally, the posterior distributions for the parameters of the reionization model (hereafter referred to as ``reionization parameters'') can be inferred from the measurements of statistical observables of the 21~cm signal --- i.e.\ informative summaries from the data scientific point of view --- with the Monte Carlo Markov Chain (MCMC) analysis that uses an explicit (e.g.\ Gaussian) likelihood approximation (see the {\tt 21CMMC} code; \citealp{2015MNRAS.449.4246G,2017MNRAS.472.2651G,Greig2018}). However, the inference using the explicit likelihood approximation in 21~cm analysis may find itself either biased if the assumption in likelihood function is not exact or the covariance matrices in the likelihood neglect the off-diagonal elements between different wavenumbers and different redshifts, or computationally too expensive if all elements of covariance matrices are properly accounted for (\citealt{zhao21cmdelfi}, hereafter referred to as \citetalias{zhao21cmdelfi}; \citealt{2023arXiv230503074P}). To address this issue of intractable likelihood, the so-called ``implicit likelihood inference'' (ILI; \citealp{alsing2018massive,alsing2019fast, 2019arXiv191013233P, Cranmer30055,tejero-cantero2020sbi}), aka ``simulation-based inference'' (SBI) or ``likelihood-free inference'' (LFI), has been recently proposed to ``learn'' the density of the likelihood or posterior directly from data, using advanced methods in deep learning, e.g.\ conditional masked autoregressive flows (CMAFs; \citealp{papamakarios2017masked}) which is a variant of normalizing flows \citep{JMLR:v22:19-1028}. 

In \citetalias{2021arXiv210503344Z}, we introduced the density estimation likelihood-free inference (DELFI) to the 21~cm analysis for the first time and performed the posterior inference of reionization parameters from the 3D tomographic 21~cm light-cone images. The 3D CNNs were adopted therein to compress the 3D 21~cm images into informative summaries. In this paper, we improve the data compressor of 3D 21~cm images and replace the 3D CNNs in \citetalias{2021arXiv210503344Z} with the solid harmonic WST\footnote{The solid harmonic WST is implemented with the {\tt Kymatio} package (\citealt{2018arXiv181211214A}; \href{https://www.kymat.io}{https://www.kymat.io}).}, but otherwise still perform the Bayesian inference of reionization parameters in the framework of DELFI. This new approach (i.e.\ solid harmonic WST with DELFI) is dubbed ``{\tt 3D ScatterNet}'' herein. We will compare the inference results using {\tt 3D ScatterNet} and those using {\tt DELFI-3D CNN}, in order to demonstrate the improvement of the data compressor. In addition, we will compare the inference results from the 21~cm images using {\tt 3D ScatterNet} and those from the 21~cm power spectrum analysis using {\tt 21cmDELFI-PS} \citepalias{zhao21cmdelfi}, both in the framework of DELFI, in order to demonstrate whether solid harmonic WST can extract more information from the 21~cm images than the power spectrum statistic. Similar to the analysis in \citetalias{zhao21cmdelfi}, realistic effects, including thermal noise and residual foreground after removal with the singular value decomposition (SVD; \citealp{stewart1993early,wolz2015foreground,masui2013measurement}), are applied to the mock observations from the SKA. 

The rest of this paper is organized as follows. The {\tt 3D ScatterNet} methodology is introduced in Section~\ref{sec:sbi}. We describe the simulations and the application of realistic effects in Section~\ref{sec:data}, present the inference results in Section~\ref{sec:results}, and make concluding remarks in Section~\ref{sec:conclusion}. Some technical details are left to Appendix~\ref{sec:l_number} (on the effect of angular frequency), Appendix~\ref{sec:comp_km} (on the effect of smoothing scales), Appendix~\ref{sec:single} (on the dependence of light-cone volumes), Appendix~\ref{sec:nde_setting} (on the network setting and sample size), and Appendix~\ref{sec:appendix_noise} (on the performance of DELFI-3DCNN with SKA noise). Some of our results were previously summarized by us in a conference paper \citep{2023arXiv230709530Z}. Our code is publicly available on \href{https://github.com/Xiaosheng-Zhao/21cmDELFI}{GitHub} {\faGithub}.

\section{The 3D ScatterNet methodology}
\label{sec:sbi}

\subsection{Solid Harmonic Wavelet Scattering Transform}
\label{sec:st}
We briefly summarize the solid harmonic WST in this subsection, following \citet{eickenberg2017solid,eickenberg2018solid}.

Solid harmonic WST convolves the original fields with a cascade of solid harmonic wavelets, performs non-linear moduli on these convolved fields, and integrates over the coordinate space. 
The solid harmonic wavelet is defined as the solid harmonic multiplying a Gaussian  
\begin{equation}
\psi_{\ell}^{m}(\mathbf{u})=\frac{1}{(\sqrt{2 \pi})^{3}} e^{-|u|^{2} / 2}|\mathbf{u}|^{\ell} Y_{\ell}^{m}\left(\frac{\mathbf{u}}{|\mathbf{u}|}\right)\,,
\label{eq:wavelet}
\end{equation}
where $|\mathbf{u}|^{\ell} Y_{\ell}^{m}\left(\frac{\mathbf{u}}{|\mathbf{u}|}\right)$ is the solid harmonic, evaluated at the coordinate $\mathbf{u}$. The Gaussian part serves as localizing the wavelet support around zero. Here we omit the further normalization factors for simplicity. In order to capture features of the field at multiple scales, the mother solid harmonic wavelet in Equation~(\ref{eq:wavelet}) is dilated at the scale\footnote{Without loss of clarity, we refer to the expression of ``at the scale $2^j$'' as ``at the scale $j$'' hereafter for simplicity.} $2^j$, i.e.\ 
\begin{equation}
\psi_{j, \ell}^{m}(\mathbf{u})=2^{-3 j} \psi_{\ell}^{m}\left(2^{-j} \mathbf{u}\right)\,.
\end{equation}
 
The Euclidean norm as the modulus operator, also dubbed ``{\it first-order} modulus coefficient'' herein, is defined by
\begin{equation}
U[j, \ell] \mathbf{d}(\mathbf{u})=\left(\sum_{m=-\ell}^{\ell}\left|\mathbf{d} * \psi_{j, \ell}^{m}(\mathbf{u})\right|^{2}\right)^{1 / 2}\,,
\label{eq:E-norm}
\end{equation}
where the field $\mathbf{d}$ is convolved (denoted by ``$*$'' in Equation~\ref{eq:E-norm}) by the dilated solid harmonic wavelets at the scale $j$ with the angular frequency band $\ell$. 
The additional rotational phase subspace information represented by $m$ is aggregated to produce coefficients that are covariant to rotation. The translational covariance is also guaranteed by the convolution operation. 

The {\it first-order} solid harmonic wavelet scattering coefficient is given by
\begin{equation}\label{eq:shw1}
S_1 \left[\mathbf{d} ; j, \ell, q \right] =\int_{\mathbb{R}^{3}}|U[j, \ell] \mathbf{d}(\mathbf{u})|^{q} d^3\mathbf{u}\,,
\end{equation}
where the modulus is raised by the power $q$, which results in the coefficients that are sensitive to the amplitude of the field, in that the small (large) value of $q$ gives more weight to the small (large) non-zero values of the integral. Note that the point-wise transformation by the power $q$ does not change the covariance property. By integrating these coefficients over the position $\mathbf{u}$, we get the coefficients that are invariant to both translation and rotation.

\begin{figure*}
    \centering
    \includegraphics[width=\textwidth]{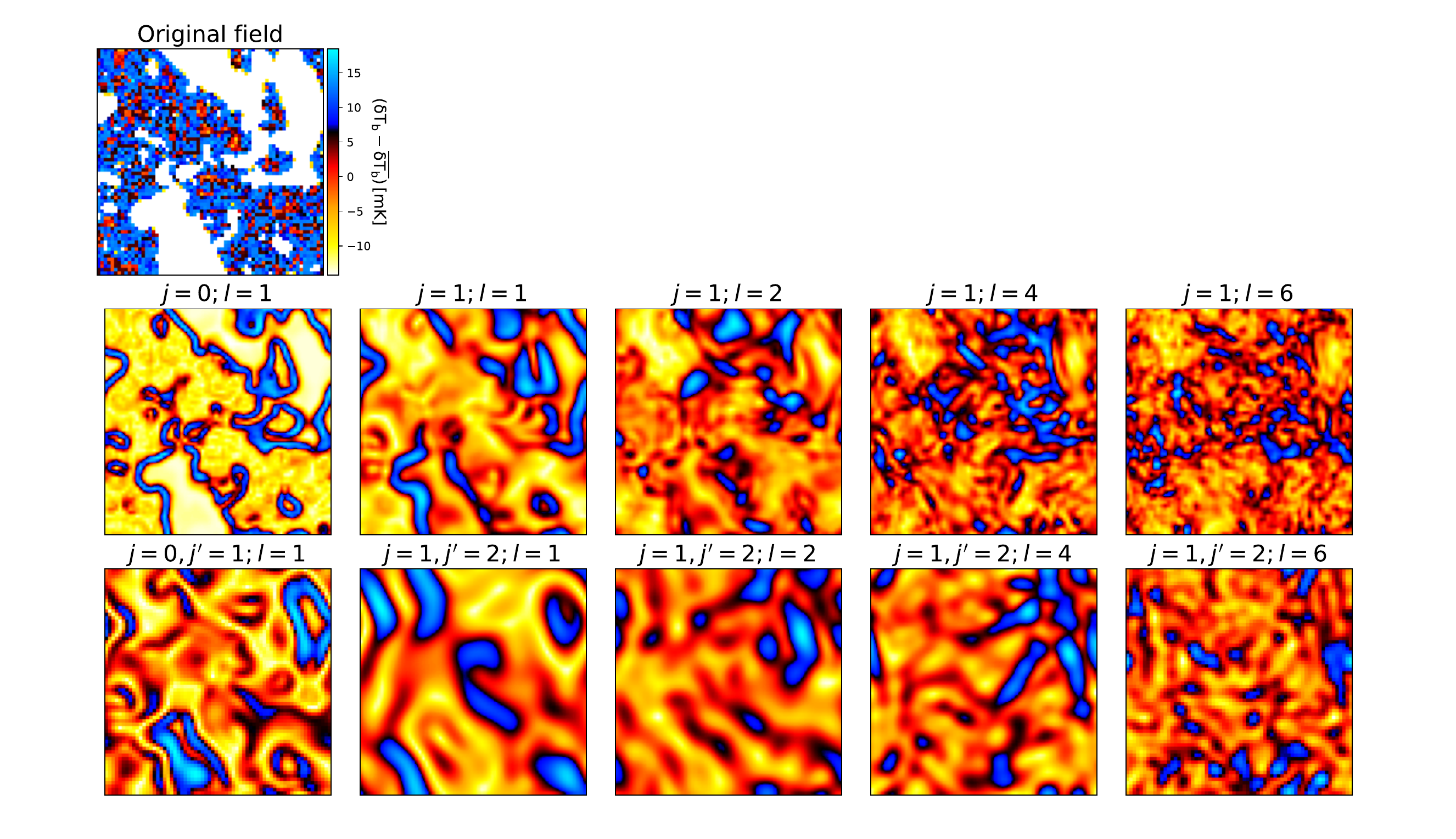}
    \caption{Visualization of solid harmonic WST. Shown is one slice of the original field of simulated cosmological 21~cm signals along the angular direction (top row), the first-order modulus coefficients $U[j, \ell] \mathbf{d}(\mathbf{u})$ (middle row) and the second-order modulus coefficients $U\left[j, j^{\prime}, \ell\right] \mathbf{d}(\mathbf{u})$ (bottom row), at some representative wavelet scales $j$ and/or $j^\prime$ and at various angular frequencies $\ell$. The colormap of all fields is the same as shown for the original field.}
    \label{fig:fields}
\end{figure*}

In order to capture the information across multiple scales, the first-order modulus coefficient $U[j, \ell] \mathbf{d}(\mathbf{u})$ is convolved with another wavelet at a different scale $j^\prime$ with $j^\prime > j $ but with the same angular frequency band $\ell$, i.e.\ the {\it second-order} modulus coefficient is defined as 
\begin{equation}
U\left[j, j^{\prime}, \ell\right] \mathbf{d}(\mathbf{u})=\left(\sum_{m=-\ell}^{\ell}\left|U[j, \ell] \mathbf{d} * \psi_{j^{\prime}, \ell}^{m}(\mathbf{u})\right|^{2}\right)^{1 / 2}, \quad j<j^{\prime}\,.
\end{equation}
The covariance property is also maintained. 

The {\it second-order} solid harmonic wavelet scattering coefficient is defined by integration over the coordinate space, similar to Equation~(\ref{eq:shw1}), 
\begin{equation}\label{eq:shw2}
S_2 \left[ \mathbf{d} ; j, j^{\prime}, \ell, q \right]=\int_{\mathbb{R}^{3}}\left|U\left[j, j^{\prime}, \ell\right] \mathbf{d}(\mathbf{u})\right|^{q} d^3 \mathbf{u}\,,
\end{equation}
which is also invariant to both translation and rotation.

In principle, successive similar operations can be applied to define the higher-order solid harmonic wavelet scattering coefficients\footnote{Without loss of clarity, we refer to the phrase ``solid harmonic wavelet scattering coefficients'' as ``scattering coefficients'' hereafter for simplicity.}. In this paper, we neglect the information encoded in the higher-order scattering coefficients and only consider the zeroth-, first- and second-order scattering coefficients, following \citet{allys2019rwst,cheng2020new}. 
In the same spirit as in Equations~(\ref{eq:shw1}) and (\ref{eq:shw2}), the {\it zeroth-order} scattering coefficient is defined as the sum of all pixel values raised by the power $q$, i.e.\ 
\begin{equation}
S_0 \left [\mathbf{d}; q\right]=\int_{\mathbb{R}^{3}}\mathbf{d}(\mathbf{u})^{q} d^3 \mathbf{u}\,. 
\end{equation}

\begin{figure*}
    \centering
    \includegraphics[width=\textwidth]{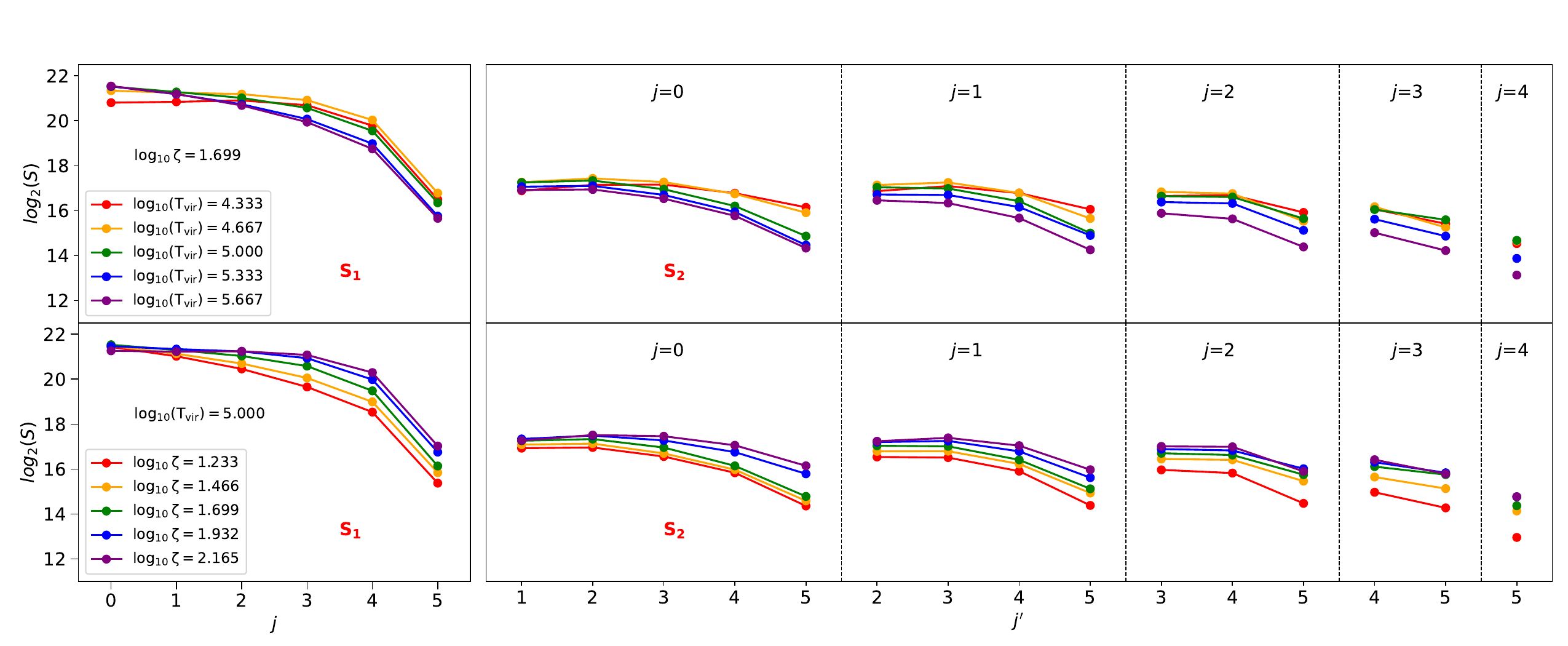}
    \caption{The first-order (``$\mathbf{S_1}$'') and second-order (``$\mathbf{S_2}$'') scattering coefficients of the simulated cosmological 21~cm signals and their dependence on the reionization parameters. The scattering coefficients are averaged over $0 \le \ell \le \ell_{\rm max}$ and evaluated at $q=1$. We vary two reionization parameters separately, namely, $\log _ { 10 } \left( T_ { \rm vir } \right)$ (top) and $\mathrm{log_{10}\zeta}$ (bottom) --- see their definitions in Section~\ref{sec:sim} ---  with the fiducial model in green lines. }
    \label{fig:statistics}
\end{figure*}

Basically, while the zeroth-order scattering coefficients highlight the amplitude of the field $\mathbf{d}$, the first-order scattering coefficients separate the information of the field into different scales by gathering the specific geometric feature in the original field at the scale $j$. Also, the second-order scattering coefficients represent the information of nonlinear mode mixing between different scales $j$ and $j^{\prime}$. These nonlinear operations obviously extract a richer set of information than the power spectrum analysis that only characterizes the features at separate Fourier modes. In addition, not only does the WST capture the local spatial information with the localized wavelets, but also it saves some large-scale information because of the long tail of the wavelets, similar to the Morlet transform \citep{10.1093/mnras/stw1310}. Furthermore, the scattering coefficients are naturally invariant to translation and rotation. Also, they are Lipschitz continuous to deformations \citep{https://doi.org/10.1002/cpa.21413}, meaning that they are approximately proportional to small deformations of the original field. 

Solid harmonic WST is very analogous to 3D CNN in three aspects as follows. Firstly, the modulus operators work similarly to the nonlinear functions in the latter. Secondly, the integration of the modulus coefficients over the coordinate space is essentially the pooling operation. Thirdly, the orders in scattering coefficients are similar to the layers in the CNN. However, solid harmonic WST has fixed kernels in wavelets, so unlike the CNN, it does {\it not} need to be trained in order to output the scattering coefficients. Also, large kernels are usually avoided in the CNN because otherwise, this would require a huge set of training parameters, but large kernels can be applied to solid harmonic WST efficiently. 

Figure~\ref{fig:fields} presents the visualization of selected modulus coefficients. Each coefficient reflects some features of the original field. For example, the first-order modulus coefficients with ($j=0$, $\ell=1$) highlight the boundaries of H~II bubbles, while the second-order modulus coefficients with ($j=0$, $ j^\prime=1$, $\ell =1$) show more complex structures. Also, as $\ell$ increases (from the left to right in Figure~\ref{fig:fields}), the modulus coefficients encode the information of small structures of the original field. Note that wavelets with $\ell=0$ are Gaussian, while wavelets with nonzero $\ell$ can encode rich structures such as filaments. 

To compute the scattering coefficients, we choose the maximum scale $j_{\rm max}=5$ and maximum angular frequency $\ell_{\rm max}=6$, and select the value of the modulus power $q = 0.5$, $1$, or $2$ for the first and second-order scattering coefficients. The half-width parameter (i.e.\ standard deviation in the Gaussian factor) of the mother solid harmonic wavelets is set to be unity, in which case the maximum scale $j_{\rm max}=5$, which satisfies the criterion $2\times {\rm (half\, width\,parameter)} \times 2^{j_{\rm max}}\le$ the number of simulation cells on each side (66 in this work, as shown in Section~\ref{sec:sim}).
For a given $\ell$ (with $0\le \ell \le \ell_{\rm max}$) and $q$, the number of the first order coefficients is $j_{\rm max}+1=6$ since $j$ takes the value of $0\le j \le j_{\rm max}$ and the number of the second order coefficients is $j_{\rm max}(j_{\rm max}+1)/2 = 15$ since $j$ and $j^\prime$ take the value of $j=0$, $1\le j^\prime \le j_{\rm max}$; $j=1$, $2\le j^\prime \le j_{\rm max}$; etc. 
To keep the dimension of coefficients reasonably low, in this paper, we simply average the information over different $\ell$ for a given $q$. (We will discuss the effect of angular frequencies $\ell$ on the parameter inference in Appendix~\ref{sec:l_number}.) Therefore, the first and second-order coefficients have a total of $63$ components for three values of power $q = 0.5$, $1$, and $2$. 

The zeroth-order coefficient is ill-defined if $q=0.5$ because the field value at a single pixel can be negative. Also, the zeroth-order coefficient is zero if $q=1$ because it is simply the global sum of the field in this case which is zero for the 21~cm signal by definition (see Section~\ref{sec:sim}). Therefore, instead of $q = 0.5$ or $1$, we choose (arbitrarily) three higher modulus power $q=2$, $3$, and $4$ in this paper for the zeroth-order coefficients. 

Altogether, the final concatenated coefficient vector has a dimension of $66$, which can be adjusted if necessary. Furthermore, following \citet{allys2019rwst}, we take the logarithms of those coefficients with base 2 as our new coefficients. (If a component is negative, we take the logarithm of the absolute value but keep the sign of the component.) The advantage of adopting these new coefficients is that they behave linearly with  $j$ and $j^\prime$. Hereafter the phrase ``scattering coefficients'' refers to their logarithms. 

We show the representative values of scattering coefficients in Figure~\ref{fig:statistics}. We vary the reionization parameters (see their definitions in Section~\ref{sec:sim}), and find that the coefficients are generically suppressed (enhanced) as $\log _ { 10 } T_ { \rm vir }$ ($\mathrm{log_{10}\zeta}$) increases. This implies that the scattering coefficients are sensitive to the reionization parameters. 

\begin{figure*}
    \centering
    \includegraphics[width=\textwidth]{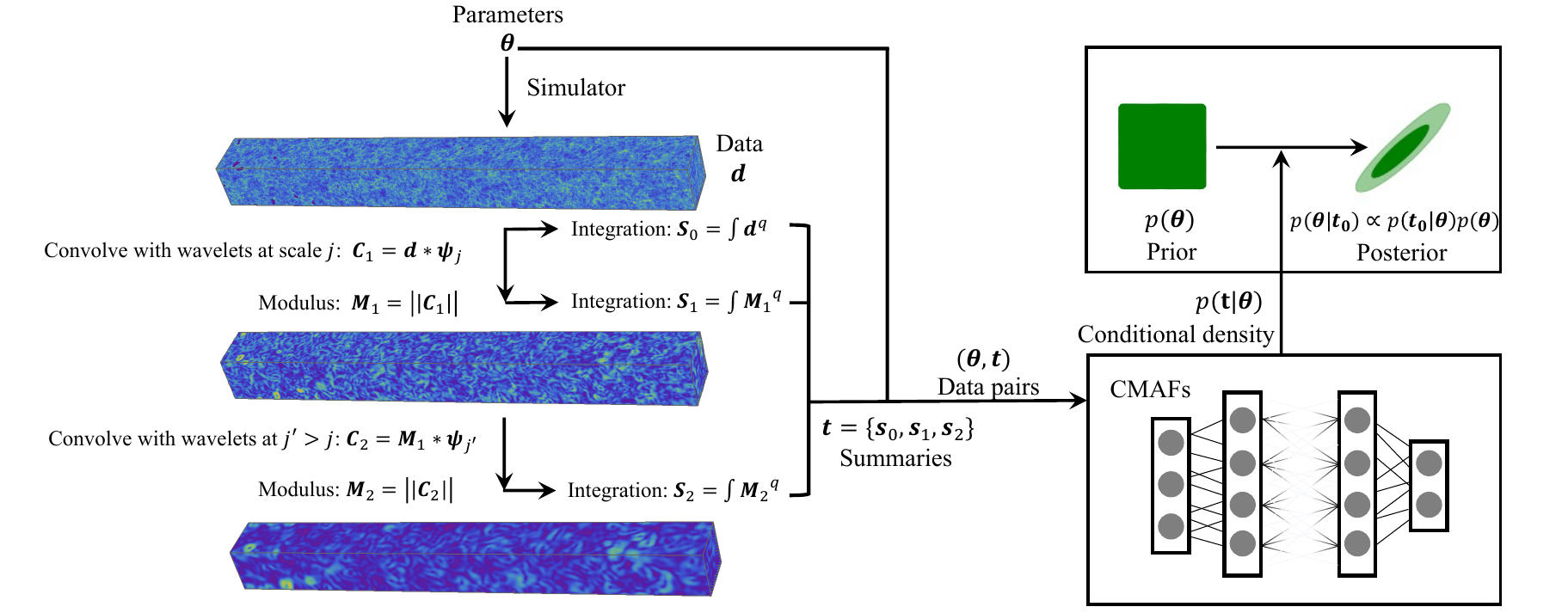}
    \caption{The workflow of {\tt 3D ScatterNet}. (Left, fixed part) the data compression with solid harmonic WST --- the 3D light-cone images $\mathbf{d}$ which are simulated with the parameters $\boldsymbol{\theta}$ are compressed by a cascade of scattering transforms (each containing the convolution with solid harmonic wavelets, harmonic modulus, and integration operation) to form the summaries $\mathbf{t}$ which include the zeroth-, first-, and second-order scattering coefficients $\{\mathbf{S_0}, \mathbf{S_1}, \mathbf{S_2}\}$, respectively. (Right, trainable part) The CMAFs are used to learn the summary density conditional on the parameters (i.e.\ likelihood). With the likelihood, the posterior can be inferred using Bayes' Theorem at the data summary $\mathbf{t_0}$.}
    \label{fig:DELFI-ST}
\end{figure*}

\subsection{Simulation-based inference with CMAFs}
\label{sec:delfi}

Figure~\ref{fig:DELFI-ST} shows the workflow of {\tt 3D ScatterNet}. The scattering coefficients extracted by the solid harmonic WST serve as the input summaries of the CMAFs\footnote{The CMAFs are implemented with {\tt pydelfi} (\citealt{alsing2019fast}; \href{https://github.com/justinalsing/pydelfi}{https://github.com/justinalsing/pydelfi}).}, which is a variant of the neural density estimators (NDEs) that perform the implicit likelihood inference. We refer interested readers to \citetalias{2021arXiv210503344Z} for an in-depth description of the DELFI and the CMAFs. The detailed settings of NDEs are given in Appendix~\ref{sec:nde_setting}. 
 In this paper, we set two neural layers of a single transform and 50 neurons per layer for CMAFs. We also use the ensembles of CMAFs to improve the performance. The number of transforms and the configuration of ensembles are chosen based on the performance of posterior validation. 

\subsection{Validation of the posterior}
\label{sec:validation_method}

As the final step of inference, we validate both marginalized and joint posteriors. 
In the posterior validation, hypothesis tests are made to check if the posteriors from CMAFs are self-consistent statistically. Note that the posterior validation is sometimes referred to as ``posterior calibration''. However, this step is not to calibrate the trained networks, but simply an indication of whether the network complexity and training data are adequate to learn the conditional density accurately in a statistical manner. We follow Appendix~A of \citetalias{2021arXiv210503344Z} and recap the validation statistics in this subsection.

For the marginalized posterior, the probability integral transform (PIT; \citealt{gneiting2007probabilistic,mucesh2021machine}) is defined as 
\begin{equation}
	{\rm PIT}\,(\tilde{\theta}) \equiv \int_{-\infty}^{\tilde{\theta}} f(\theta) \,\mathrm{d} \theta\,,
\end{equation}
i.e.\ the cumulative distribution function (CDF) of the inferred marginal distribution $f(\theta)$ at the true value $\tilde{\theta}$. If the inferred posteriors are accurate, then the PIT should be uniformly distributed.

For the joint posteriors, there are two statistics. The first one is the copula probability integral transformation (copPIT; \citealt{10.1214/14-EJS964,mucesh2021machine}) which is defined as 
\begin{equation}
	{\rm copPIT} \equiv \operatorname{Pr}\{H(\boldsymbol{\theta}) \leq H(\tilde{\boldsymbol{\theta}})\}\,. 
	\end{equation}
Here ``Pr'' represents the probability and $H(\boldsymbol{\theta})$ is the CDF of the inferred joint distribution. The copPIT is the multivariate extension of PIT. 

The second statistic is the highest probability density (HPD; \citealt{harrison2015validation}) which is defined as
\begin{equation}
	{\rm HPD}\,(\boldsymbol{\tilde{\theta}}) \equiv \int_{f(\boldsymbol{\theta}) \geq f(\boldsymbol{\tilde{\theta}})} f(\boldsymbol{\theta}) \mathrm{d}^{n} \boldsymbol{\theta}\,.
	\end{equation}
The HPD describes the plausibility of $\boldsymbol{\tilde{\theta}}$ under the distribution $f(\boldsymbol{\theta})$. A small value of HPD indicates high plausibility. Similar to the PIT, the copPIT and HPD should be also uniformly distributed if the posteriors are accurate. 

In order to check the uniformity, we adopt two metrics, the Kolmogorov-Smirnov (KS; \citealp{Kolmogorov1992}) test and Cram\'er-von Mises {(CvM; \citealp{10.1214/aoms/1177704477})} test, which focus on different aspects of distribution. While the KS test is sensitive to the median, the CvM test captures the tails of a distribution. If the $p$-value from a test is larger than the preset value of significance level, typically 0.01 or 0.05 \citep{Ivezic2014}, then the null hypothesis that these statistics follow a uniform distribution is accepted. 
Throughout this paper, we adopt the significance level of 0.01 and report the results only if the $p$-value is larger than 0.01 unless stated otherwise. 

\section{Data preparation}
\label{sec:data}

\begin{figure*}
    \centering
    \includegraphics[width=\textwidth]{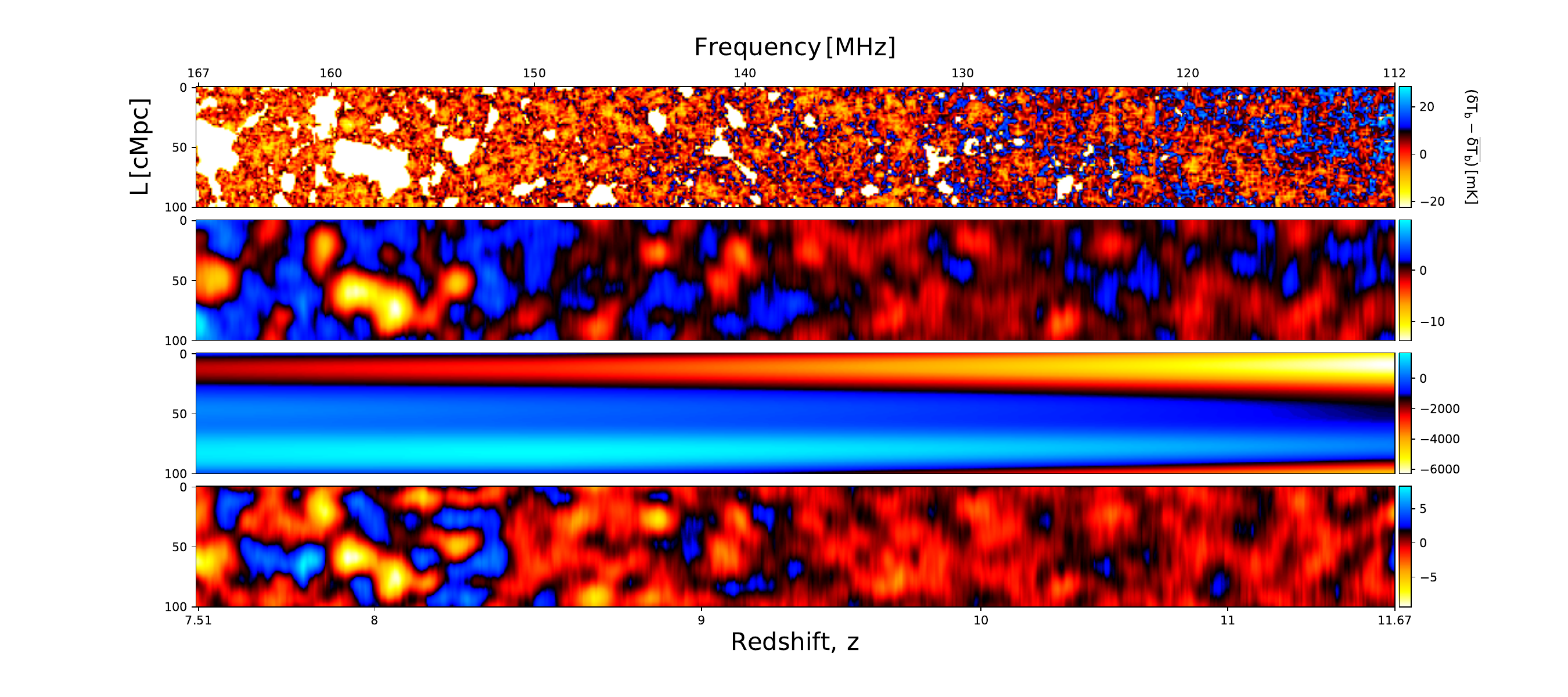}
    \caption{Visualization of the mock 21~cm observation in a slice along the LOS. From top to bottom: (1) the simulated cosmic 21~cm images without realistic effect applied except that the ${\bf k}_\perp = 0$ mode is removed from each 2D slice perpendicular to the LOS (i.e.\ ``pure signal''); (2) same as the pure signal but with total noise from SKA observation (i.e.\ ``SKA noise''); (3) same as the pure signal but with total noise from SKA observation and the foreground contamination; (4) same as the pure signal but with total noise from SKA observation and residual foreground after the foreground is removed with the SVD technique (i.e.\ ``SKA noise + residual foreground''). The map of pure signal retains the original grid size in simulations, but the rest of the three maps are made by convolving the original maps with a Gaussian filter in the angular direction and a top-hat filter in the LOS direction, with the widths of the filters corresponding to the size of the 1-km baseline.} 
    \label{fig:instru}
\end{figure*}

\subsection{Cosmic 21 cm Signal}
\label{sec:sim}
The 21~cm brightness temperature relative to the CMB temperature at position ${\bf x}$ can be written \citep{Furlanetto2006} as 
\begin{equation}
T_{21}(\textbf{x},z)=\tilde{T}_{21}(z)\,x_{\rm HI}(\textbf{x})\,\left[1+\delta(\textbf{x})\right]\,\left(1-\frac{T_{\rm CMB}}{T_S}\right)\,,\label{eqn:21cm}
\end{equation}
where $\tilde{T}_{21}(z) = 27\sqrt{[(1+z)/10](0.15/\Omega_{\rm m} h^2)}(\Omega_{\rm b} h^2/0.023)$ in units of mK, $x_{\rm HI}({\bf x})$ is the neutral fraction, and $\delta({\bf x})$ is the matter overdensity, at position ${\bf x}$. We assume the cold
dark matter can be traced by the baryon perturbation on large scales, so $\delta_{\rho_{\rm H}} = \delta$. 
In this paper, we focus on the limit where spin temperature $T_S \gg T_{\rm CMB}$, likely valid soon after reionization begins, though this assumption is strongly model-dependent. As such, we can neglect the dependence on spin temperature. Also, as a demonstration of concept, we ignore the effect of peculiar velocity; such an effect can be readily incorporated in forward simulations by the algorithm introduced by \cite{2012MNRAS.422..926M}. 

In this paper, we use the publicly available code \mbox{{\tt 21cmFAST}}\footnote{\href{https://github.com/andreimesinger/21cmFAST}{https://github.com/andreimesinger/21cmFAST}} \citep{Mesinger2007,Mesinger2011}, which can be used to perform semi-numerical simulations of reionization, as the simulator to generate the datasets. Our simulations were performed on a cubic box of 100 comoving Mpc on each side, with $66^3$ grid cells. Following the interpolation approach in \citetalias{2021arXiv210503344Z}, the snapshots at nine different redshifts of the same simulation box (i.e., with the same initial condition) are interpolated to construct a light-cone 21~cm data cube within the comoving distance of a simulation box along the line of sight (LOS). We concatenate 10 such light-cone boxes, each simulated with different initial conditions in density fields but with the same reionization parameters, together to form a full light-cone datacube\footnote{We investigate the contributions of individual waveband to parameter inference in Appendix~\ref{sec:single}.} of the size $100\times 100 \times 1000$ comoving ${\rm Mpc}^3$ (or $66\times 66 \times 660$ grid cells) in the redshift range $7.51 \le z \le 11.67$. To mimic the observations from radio interferometers, we subtract from the light-cone field the mean of the 2D slice for each 2D slice perpendicular to the LOS, because radio interferometers cannot measure the mode with ${\bf k}_\perp = 0$. 

We parametrize our reionization model as follows, and refer interested readers to \citetalias{2021arXiv210503344Z} for a detailed explanation of their physical meanings.

(1) $\zeta$, the ionizing efficiency, which is a combination of several parameters related to ionizing photons. In our paper, we vary $\zeta$ as $10 \le \zeta \le 250 $.

(2) $T_ { \mathrm { vir } }$, the minimum virial temperature of halos that host ionizing sources. In our paper, we vary this parameter as $ 4 \le \log _ { 10 } \left( T_{ \mathrm { vir } } / \mathrm { K } \right) \le 6 $.

Cosmological parameters are fixed in this paper as ($\mathrm{ \Omega_{\Lambda}}$, $\mathrm{\Omega _ { m }}$, $\mathrm{\Omega _ { b }}$, $\mathrm{n_s}$, $\mathrm{\sigma _ { 8 }}$, $h$) = (0.692, 0.308, 0.0484, 0.968, 0.815, 0.678) \citep{ade2016planck}. 

\subsection{Thermal noise and residual foreground}
\label{sec:thermal noise}

To apply the realistic effects to the 21~cm signal, we first compute the $uv$ coverage map at each of a total of 660 frequency channels. The $uv$ coverage map can be used to suppress the thermal noise and calculate the telescope response to the cosmic 21 cm signal and foreground. We then generate the thermal noise with a 1000-hour observation. For the thermal noise estimation in this paper, we employ the {\tt Tools21cm}\footnote{\href{https://github.com/sambit-giri/tools21cm}{https://github.com/sambit-giri/tools21cm}} \citep{Giri2020} code to simulate the expected thermal noise in the 3D 21~cm light-cone images as observed with the SKA1-Low. We list our basic assumptions of the SKA configuration in Table~\ref{tab:instru parameters}.  
When synthesizing the total signal, we applied the same telescope response in the calculation of thermal noise to both the cosmic 21~cm signal component and the radio foreground component. The signal of a pixel in the $uv$ space is kept unchanged for non-zero $uv$ coverage but otherwise set to zero if the pixel is outside the $uv$ coverage. In order to suppress the noise, we smooth the map with a relatively small baseline of 1~km\footnote{\citet{2018MNRAS.479.5596G,giri2020measuring,10.1093/mnras/stab1518} choose to smooth with a baseline of 2~km, which corresponds to the size of the central area of SKA1-Low. We discuss the effect of smoothing scales in Appendix~\ref{sec:comp_km}.}, which roughly corresponds to the size of the core area of SKA1-Low. Specifically, we convolve the map with a Gaussian filter in the angular direction and a top-hat filter in the LOS direction. The FWHM of the Gaussian filter corresponds to the size of the 1-km baseline and the width of the top-hat filter also matches that FWHM. 

\begin{table}
	\centering
	\caption{Specifications for the SKA}
	\begin{tabularx}{\linewidth}{lr} 
		\hline
		Parameters    & Values \\
		\hline
        System temperature  & $60(\frac{v}{300 \mathrm{MH}})^{-2.55} + 100 \mathrm{~K}$ \\ 
		Effective collecting area             & $962\ \mathrm{m^2}$  \\
		Integration time               & 10 seconds \\
		Observation hours per day   & 6 hours \\
		Total observation time           & 1000 hours \\
		\hline
	\end{tabularx}
	\label{tab:instru parameters}
\end{table}

To mock the foreground contamination, we employ the {\tt pygsm}\footnote{\href{https://github.com/telegraphic/pygsm}{https://github.com/telegraphic/pygsm}} package that is based on the GSM-building model \citep{zheng2017improved}, which interpolates the sky maps with 29 sky map observations using an improved principal component analysis (PCA) method. We then use these foreground maps to interpolate our images on the grid for each frequency slice. In order to prevent overfitting, we assign a random patch of sky (except for the North and South poles) to model each foreground image. Finally, we employ the SVD technique for foreground removal. Specifically, we remove the largest six singular value modes. 

For visualization purposes, we show the mock 21~cm observation with realistic effects in Figure~\ref{fig:instru}. When the cosmic 21~cm signal is applied with the thermal noise of SKA and/or the residual foreground, the features at small scales are washed out but those at the very large scales are still retained. 

\begin{table*}
	\caption{Bayesian Inference with {\tt 21cmDELFI-PS}, {\tt DELFI-3D CNN} and {\tt 3D ScatterNet} for the ``Faint Galaxies Model''}
	\label{tab: stat_a}
		\begin{tabular*}{\textwidth}{l @{\extracolsep{\fill}} *{8}{c}}
			\hline\hline
					 & &\multicolumn{3}{c}{\scriptsize{Pure signal}} &\multicolumn{2}{c}{\scriptsize{SKA noise}} &\multicolumn{2}{c}{\scriptsize{SKA noise + residual foreground}} \\
					 \cmidrule(l{.75em}r{.75em}){3-5}
					 \cmidrule(l{.75em}r{.75em}){6-7}
					 \cmidrule(l{.75em}r{.75em}){8-9}
			\scriptsize{Parameter} & \scriptsize{True value} & \scriptsize{{\tt DELFI-3DCNN}} & \scriptsize{ {\tt 21cmDELFI-PS}} & \scriptsize{ {\tt 3D ScatterNet}} & \scriptsize{ {\tt 21cmDELFI-PS}} &  \scriptsize{ {\tt 3D ScatterNet}} & \scriptsize{ {\tt 21cmDELFI-PS}} & \scriptsize{ {\tt 3D ScatterNet}}\\
				 \hline
				 \scriptsize{$\log _ { 10 } \left( T_ { \rm vir }/{\rm K}\right)$} & \scriptsize{4.699} & \scriptsize{$4.697^{+0.024}_{-0.024} $} & \scriptsize{$4.699^{+0.011}_{-0.011}  $} & \scriptsize{$4.701^{+0.006}_{-0.006}$} & \scriptsize{$4.716^{+0.081}_{-0.070} $} & \scriptsize{$4.713^{+0.085}_{-0.059}$} & \scriptsize{$4.789^{+0.185}_{-0.152}$} & \scriptsize{$4.774^{+0.133}_{-0.120}$}\\
				 \hline
		   \scriptsize{$\log _ { 10 }(\zeta)$} & \scriptsize{1.477} & \scriptsize{$1.475^{+0.023}_{-0.023}$} & \scriptsize{$1.481^{+0.011}_{-0.011}$} & \scriptsize{$1.479^{+0.007}_{-0.006}$} & \scriptsize{$1.508^{+0.051}_{-0.043} $} & \scriptsize{$1.498^{+0.059}_{-0.038}$} & \scriptsize{$1.552^{+0.123}_{-0.099}$} & \scriptsize{$1.533^{+0.097}_{-0.077}$}\\
		\hline
		\end{tabular*}
		\flushleft
		\tablenotetext{}{{\bf Note.} Here, ``pure signal'' refers to the mock observations of cosmological 3D 21~cm images (i.e.\ without thermal noise or foreground contamination, but the ${\bf k}_\perp = 0$ mode is removed from each 2D slice perpendicular to the LOS);  ``SKA noise'' refers to the mock SKA observations of the 3D 21~cm images with total noise (with the contributions from thermal noise and sample variance errors) yet without foreground contamination; ``SKA noise + residual foreground'') refers to the mock SKA observations of the 3D 21~cm images with total noise and residual foreground after the foreground is removed with the SVD technique.}
\end{table*}

\begin{table*}
	\caption{Same as Table~\ref{tab: stat_a} but for the ``Bright Galaxies Model''}
	\label{tab: stat_b}
		\begin{tabular*}{\textwidth}{c @{\extracolsep{\fill}} *{8}{c}}
			\hline\hline
					 & &\multicolumn{3}{c}{\scriptsize{Pure signal}} &\multicolumn{2}{c}{\scriptsize{SKA noise}} &\multicolumn{2}{c}{\scriptsize{SKA noise + residual foreground}} \\
					 \cmidrule(l{.75em}r{.75em}){3-5}
					 \cmidrule(l{.75em}r{.75em}){6-7}
					 \cmidrule(l{.75em}r{.75em}){8-9}
			\scriptsize{Parameter} & \scriptsize{True value} & \scriptsize{{\tt DELFI-3DCNN}} & \scriptsize{ {\tt 21cmDELFI-PS}} & \scriptsize{ {\tt 3D ScatterNet}} & \scriptsize{ {\tt 21cmDELFI-PS}} &  \scriptsize{ {\tt 3D ScatterNet}} & \scriptsize{ {\tt 21cmDELFI-PS}} & \scriptsize{ {\tt 3D ScatterNet}}\\
				 \hline
				 \scriptsize{$\log _ { 10 } \left( T_ { \rm vir }/{\rm K}\right)$} & \scriptsize{5.477} & \scriptsize{$5.485^{+0.037}_{-0.036}$} & \scriptsize{$5.480^{+0.015}_{-0.016} $} & \scriptsize{$5.476^{+0.010}_{-0.010}$} & \scriptsize{$5.464^{+0.063}_{-0.088} $} & \scriptsize{$5.446^{+0.060}_{-0.068}$} & \scriptsize{$5.341^{+0.128}_{-0.139}$} & \scriptsize{$5.349^{+0.132}_{-0.137}$} \\
				 \hline
		   \scriptsize{$\log _ { 10 }(\zeta)$} & \scriptsize{2.301} & \scriptsize{$2.307^{+0.036}_{-0.033}$} & \scriptsize{$2.306^{+0.023}_{-0.023}$} & \scriptsize{$2.308^{+0.021}_{-0.019}$} & \scriptsize{$2.279^{+0.078}_{-0.110} $} & \scriptsize{$2.258^{+0.077}_{-0.085}$} & \scriptsize{$2.117^{+0.149}_{-0.155}$} & \scriptsize{$2.122^{+0.164}_{-0.154}$} \\
		\hline
		\end{tabular*}
\end{table*}

\begin{table*}
		\caption{Recovery Performance by the {\tt 21cmDELFI-PS} and {\tt 3D ScatterNet}} 
		\label{tab:perf_comparison}

		\begin{tabular*}{\textwidth}{c @{\extracolsep{\fill}} *{7}{c}}
			\hline \hline
			& &\multicolumn{2}{c}{\scriptsize{Pure signal}} &\multicolumn{2}{c}{\scriptsize{SKA noise}} &\multicolumn{2}{c}{\scriptsize{SKA noise + residual foreground}} \\
					 \cmidrule(l{.75em}r{.75em}){3-4}
					 \cmidrule(l{.75em}r{.75em}){5-6}
					 \cmidrule(l{.75em}r{.75em}){7-8}
		  {}  & {} & \scriptsize{ {\tt 21cmDELFI-PS}} & \scriptsize{ {\tt 3D ScatterNet}} & \scriptsize{ {\tt 21cmDELFI-PS}} &  \scriptsize{ {\tt 3D ScatterNet}} & \scriptsize{ {\tt 21cmDELFI-PS}} & \scriptsize{ {\tt 3D ScatterNet}}\\
				\hline
		\multirow{2}{*}{\scriptsize{$R^2$}\,\tablenotemark{\scriptsize{a}}} & \scriptsize{$ \log_{10}(T_{\rm vir})$} & \scriptsize{0.9989} & \scriptsize{0.9997} & \scriptsize{0.9336} & \scriptsize{0.9647} & \scriptsize{0.7981} & \scriptsize{0.8348} \\ 
		{} & \scriptsize{$\log_{10}(\zeta)$} & \scriptsize{0.9978} & \scriptsize{0.9990} & \scriptsize{0.9353} & \scriptsize{0.9604} & \scriptsize{0.8028} & \scriptsize{0.8254}\\
			\hline
		 \multirow{2}{*}{\scriptsize{$\epsilon$} \,\tablenotemark{\scriptsize{b}}} & \scriptsize{$ T_{\rm vir}$} & \scriptsize{$(-0.03, 0.03)$} & \scriptsize{$(-0.02, 0.02)$} & \scriptsize{$(-0.19, 0.30)$} & \scriptsize{$(-0.17, 0.16)$} & \scriptsize{$(-0.28, 0.47)$} & \scriptsize{$(-0.28, 0.41)$}  \\ 
		{} & \scriptsize{$\zeta$} & \scriptsize{$(-0.04, 0.03)$} & \scriptsize{$(-0.02, 0.02)$} & \scriptsize{$(-0.15, 0.24)$} & \scriptsize{$(-0.12, 0.14)$} & \scriptsize{$(-0.26, 0.36)$} & \scriptsize{$(-0.22, 0.32)$} \\
		\hline 
		\end{tabular*}
		\flushleft
			\tablenotetext{\scriptsize{a}}{The coefficient of determination $R^2$ is computed for the medians (in the logarithmic scale) of the inferred posteriors from 300 testing samples.}
			\tablenotetext{\scriptsize{b}}{The fractional error $\epsilon$ refers to the relative error of the deduced parameters in the linear scale. Here we present the $68\%$ credible interval of the probability density distribution of $\epsilon$.}
\end{table*}

\begin{table*}
		\caption{The $p$-values for the Null Hypotheses That These Statistics Are of a Uniform Distribution with 300 Testing Samples}
		\label{tab: bayesian num}
		\begin{tabular*}{\textwidth}{c @{\extracolsep{\fill}} *{12}{c}}		
			\hline\hline
			 &\multicolumn{4}{c}{\scriptsize{Pure signal}}&\multicolumn{4}{c}{\scriptsize{SKA noise}}&\multicolumn{4}{c}{\scriptsize{SKA noise + residual foreground}}\\
			 \cmidrule(l{.75em}l{.75em}r{.75em}){2-5}
			 \cmidrule(l{.75em}l{.75em}r{.75em}){6-9}
			 \cmidrule(l{.75em}l{.75em}r{.75em}){10-13}
             &\multicolumn{2}{c}{\scriptsize{{\tt 21cmDELFI-PS}}}&\multicolumn{2}{c}{\scriptsize{{\tt 3D ScatterNet}}}&\multicolumn{2}{c}{\scriptsize{{\tt 21cmDELFI-PS}}}&\multicolumn{2}{c}{\scriptsize{{\tt 3D ScatterNet}}}&\multicolumn{2}{c}{\scriptsize{{\tt 21cmDELFI-PS}}}&\multicolumn{2}{c}{\scriptsize{{\tt 3D ScatterNet}}}\\
             \cmidrule(l{.75em}l{.75em}r{.75em}){2-3}
			 \cmidrule(l{.75em}l{.75em}r{.75em}){4-5}
			 \cmidrule(l{.75em}l{.75em}r{.75em}){6-7}  
             \cmidrule(l{.75em}l{.75em}r{.75em}){8-9} 
             \cmidrule(l{.75em}l{.75em}r{.75em}){10-11}  
             \cmidrule(l{.75em}l{.75em}r{.75em}){12-13}      
			\scriptsize{Statistics} &\scriptsize{KS} & \scriptsize{CoM} & \scriptsize{KS} & \scriptsize{CoM} & \scriptsize{KS} & \scriptsize{CoM} & \scriptsize{KS} & \scriptsize{CoM} & \scriptsize{KS} &\scriptsize{CoM} & \scriptsize{KS} & \scriptsize{CoM} \\
			\hline
			\scriptsize{PIT ($T_{\rm vir}$)}   & \scriptsize{0.37} & \scriptsize{0.13} & \scriptsize{0.80} & \scriptsize{0.65}    & \scriptsize{0.01} & \scriptsize{0.02}& \scriptsize{0.20} & \scriptsize{0.52} & \scriptsize{0.19} & \scriptsize{0.17} & \scriptsize{0.42} & \scriptsize{0.37} \\
			\hline
			\scriptsize{PIT ($\zeta$)} & \scriptsize{0.14}  & \scriptsize{0.05} & \scriptsize{0.92}  &\scriptsize{0.75}    & \scriptsize{0.05} & \scriptsize{0.05} & \scriptsize{0.78} & \scriptsize{0.65} & \scriptsize{0.01} & \scriptsize{0.02} & \scriptsize{0.18} & \scriptsize{0.24} \\
			\hline
			\scriptsize{copPIT} & \scriptsize{0.54}  & \scriptsize{0.39} & \scriptsize{0.67}   & \scriptsize{0.92}     & \scriptsize{0.01}  & \scriptsize{0.03}  & \scriptsize{0.64} & \scriptsize{0.85} & \scriptsize{0.53}  & \scriptsize{0.31} & \scriptsize{0.46}  & \scriptsize{0.46} \\
			\hline
			\scriptsize{HPD}  & \scriptsize{0.04}   & \scriptsize{0.01}  & \scriptsize{0.63}   &\scriptsize{0.48}     & \scriptsize{0.01}  & \scriptsize{0.01} & \scriptsize{0.41} & \scriptsize{0.47} & \scriptsize{0.74}  & \scriptsize{0.67} & \scriptsize{0.05}  & \scriptsize{0.03} \\
			\hline
		\end{tabular*}
\end{table*}

\begin{figure*}
    \centering
    \includegraphics[width=\textwidth]{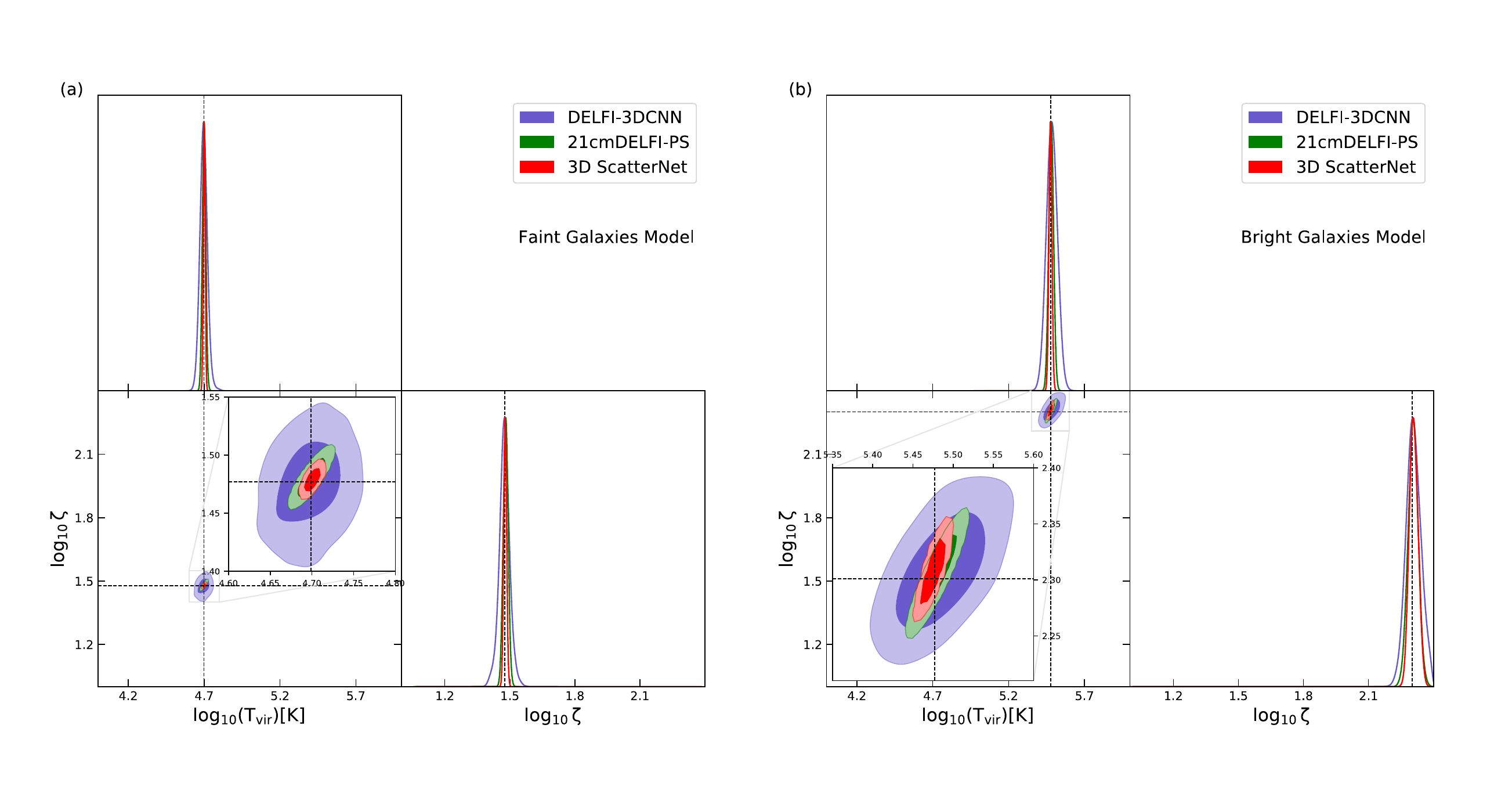}
    \caption{The posteriors estimated from 3D cosmic 21~cm light-cone images (i.e., without thermal noise or residual foreground) by three different approaches of data compression --- power spectrum using {\tt 21cmDELFI-PS} (green), 3D CNN using {\tt DELFI-3D CNN} (blue) and solid harmonic WST using {\tt 3D ScatterNet} (red) --- for two mock observations: the ``Faint Galaxies Model'' (left) and the ``Bright Galaxies Model'' (right). We show the $1\sigma$ (dark) and $2\sigma$ (light) credible regions. The dashed lines indicate the true parameter values.}
    \label{fig:comp_pure}
\end{figure*}

\section{Results}
\label{sec:results}

For each approach of data compression, we first perform the inference for two representative mock observations --- the ``Faint Galaxies Model'' and the ``Bright Galaxies Model'', defined in Tables~\ref{tab: stat_a} and \ref{tab: stat_b} (see the ``True value'' therein), respectively, following the convention of \citet{2017MNRAS.472.2651G} (see, also, \citetalias{2021arXiv210503344Z} and \citetalias{zhao21cmdelfi}). They are selected as two illustrative examples with extreme parameter values which are nevertheless fine-tuned for these models to have similar global reionization histories. 
We further perform the posterior validation of the trained NDEs using another set of 300 independent samples that are randomly drawn from the allowed region in the parameter space in which the mean neutral fraction satisfies $0.08 \le \bar{x}_{\rm HI} \le 0.81$ at $z=7.1$, corresponding to the $2 \sigma$ credible region as constrained by the IGM damping wing of ULASJ1120+0641 \citep{2017MNRAS.466.4239G}. 

\subsection{3D ScatterNet vs 3D CNN}
\label{sec:results_pure}

We compare the results using {\tt 3D ScatterNet} and using {\tt DELFI-3D CNN} in this subsection. For simplicity, here we only consider the case of cosmological 3D 21~cm images without realistic effect applied except that the ${\bf k}_\perp = 0$ mode is removed from each 2D slice perpendicular to the LOS (i.e.\ ``pure signal''). Figure~\ref{fig:comp_pure} shows the credible regions ($1\sigma$ and $2\sigma$, or 0.68 and 0.95 levels) for the ``Faint Galaxies Model'' and the ``Bright Galaxies Model'', with quantitative comparisons in terms of medians and $1\sigma$ (16th and 84th percentile) errors presented in Tables~\ref{tab: stat_a} and \ref{tab: stat_b}. 

The results from {\tt DELFI-3DCNN} were taken from \citetalias{2021arXiv210503344Z}, in which the ILI was performed with data compression made by a trained 3D CNN. \citetalias{2021arXiv210503344Z} used a set of 9,000 samples for training and validation of the 3D CNN and another set of 9,000 samples for training and validation of the density estimators. Nevertheless, if the training sample size was doubled, the results for the ``Faint Galaxies Model'' and the ``Bright Galaxies Model'' were not improved, which implies that the inference accuracy might be limited by some intrinsic properties in experimental choices (including the network architecture, characteristics in training datasets, and other hyper-parameter choices) for training the 3D CNN. However, when we replace the data compression from 3D CNN to solid harmonic WST, we find that the inference results are significantly improved in terms of the location and size of credible regions in the posterior distributions for the reionization parameters, as shown in Figure~\ref{fig:comp_pure} and Tables~\ref{tab: stat_a} and \ref{tab: stat_b}. Also, the degeneracy of these two parameters is clearly revealed in the credible region for {\tt 3D ScatterNet}, which indicates that this degeneracy is intrinsic in the theoretical modeling. 

Now that the solid harmonic WST represents a better approach to data compression than 3D CNN, we will focus on the {\tt 3D ScatterNet} in the remainder of this paper. 

\subsection{3D ScatterNet vs 21cmDELFI-PS} 

\citetalias{zhao21cmdelfi} shows that for the power spectrum analysis, the DELFI approach (i.e.\ {\tt 21cmDELFI-PS}) outperforms the standard MCMC analysis. In this subsection, therefore, we compare the results using {\tt 3D ScatterNet} and {\tt 21cmDELFI-PS}, both under the same inference strategy using DELFI. On this same footing, we can investigate whether data compression using solid harmonic WST extracts more information from the 3D 21~cm light-cone images than the power spectrum statistic. 

\subsubsection{Pure Signal}

We first consider the ``pure signal'' case for the comparison, with the credible regions for the ``Faint Galaxies Model'' and the ``Bright Galaxies Model'' shown in Figure~\ref{fig:comp_pure} and quantitative results listed in Tables~\ref{tab: stat_a} and \ref{tab: stat_b}. For the ``Faint Galaxies Model'', the systematic shift (i.e.\ relative errors of the predicted medians with respect to the true values) and the $1\sigma$ statistical errors are $0.04\% \pm 0.13\%$ ($0\%\pm 0.23\%$) for $\log _ { 10 } \left( T_ { \rm vir } \right)$ with {\tt 3D ScatterNet} ({\tt 21cmDELFI-PS}), respectively\footnote{Here for this mock observation, the predicted median with {\tt 3D ScatterNet} has slightly more deviation (but still within the $1\sigma$ credible region) from the true value than that with {\tt 21cmDELFI-PS}. However, testing the medians with 300 samples, we find that the predicted medians using {\tt 3D ScatterNet} are statistically closer to the true values than those using {\tt 21cmDELFI-PS}.}, and $0.14\%{+0.47\% \atop -0.41\%}$ ($0.27\%\pm 0.74\%$) for $\mathrm{log_{10}\zeta}$ with {\tt 3D ScatterNet} ({\tt 21cmDELFI-PS}), respectively. The estimated statistical errors using {\tt 3D ScatterNet} are about $1.7$ times smaller than using {\tt 21cmDELFI-PS}. The similar results hold generically for the ``Bright Galaxies Model'', too. 

Next, we test the trained NDEs on 300 samples. Table~\ref{tab:perf_comparison} shows the coefficient of determination, $
	R  ^ { 2 }  = 1 - \sum \left( y _ { \mathrm { pred } } - y _ { \mathrm { true } } \right) ^ { 2 } /  \sum \left( y _ { \mathrm { true } } - \overline { y } _ { \mathrm { true } } \right) ^ { 2 } \,,$
where $y _ { \mathrm { true } }$ and $y _ { \mathrm { pred } }$ are the true value and the predicted median in a sample for the parameter $y$ (e.g.\ $\log _ { 10 } \left( T_ { \rm vir } \right)$ and $\mathrm{log_{10}\zeta}$ in this paper), respectively, and the summation is over all testing samples. $\overline { y } _ { \mathrm { true }}$ is the average of the true value over all testing samples. A score of $ R  ^ { \mathrm { 2 } }$ close to unity indicates an overall good inference performance of this parameter. For the case of ``pure signal'', both {\tt 3D ScatterNet} and {\tt 21cmDELFI-PS} give very high $R  ^ { 2 }$ score, but the {\tt 3D ScatterNet} slightly outperforms the {\tt 21cmDELFI-PS}. 

Table~\ref{tab:perf_comparison} also presents the $68\%$ credible interval of the probability density distribution of $\epsilon$, the fractional errors of the deduced parameter $x$ in the linear scale (e.g.\ $ T_{\rm vir}$ and $\zeta$ in this paper). The fractional error refers to that of the predicted median with respect to the true value, i.e.\ $\epsilon = (  x _ { \mathrm { pred } } - x _ { \mathrm { true } }) / x _ { \mathrm { true } }$. For the case of ``pure signal'', the typical fractional error (represented by the length of the interval) of {\tt 3D ScatterNet} is about 1.50 times smaller than that of  {\tt 21cmDELFI-PS} for $T_{\rm vir}$, and 1.75 times smaller for $\zeta$. 

Lastly, we perform the posterior validation. Table~\ref{tab: bayesian num} shows the $p$-value for some hypothesis tests using 300 testing samples. Note that all $p$-values are larger than 0.01 and most are larger than 0.5, which implies that our results are at least reliable with a significance of 0.01. Also, the $p$-values of {\tt 3D ScatterNet} are generically larger than those of {\tt 21cmDELFI-PS}. 

\begin{figure*}
    \centering
    \includegraphics[width=\textwidth]{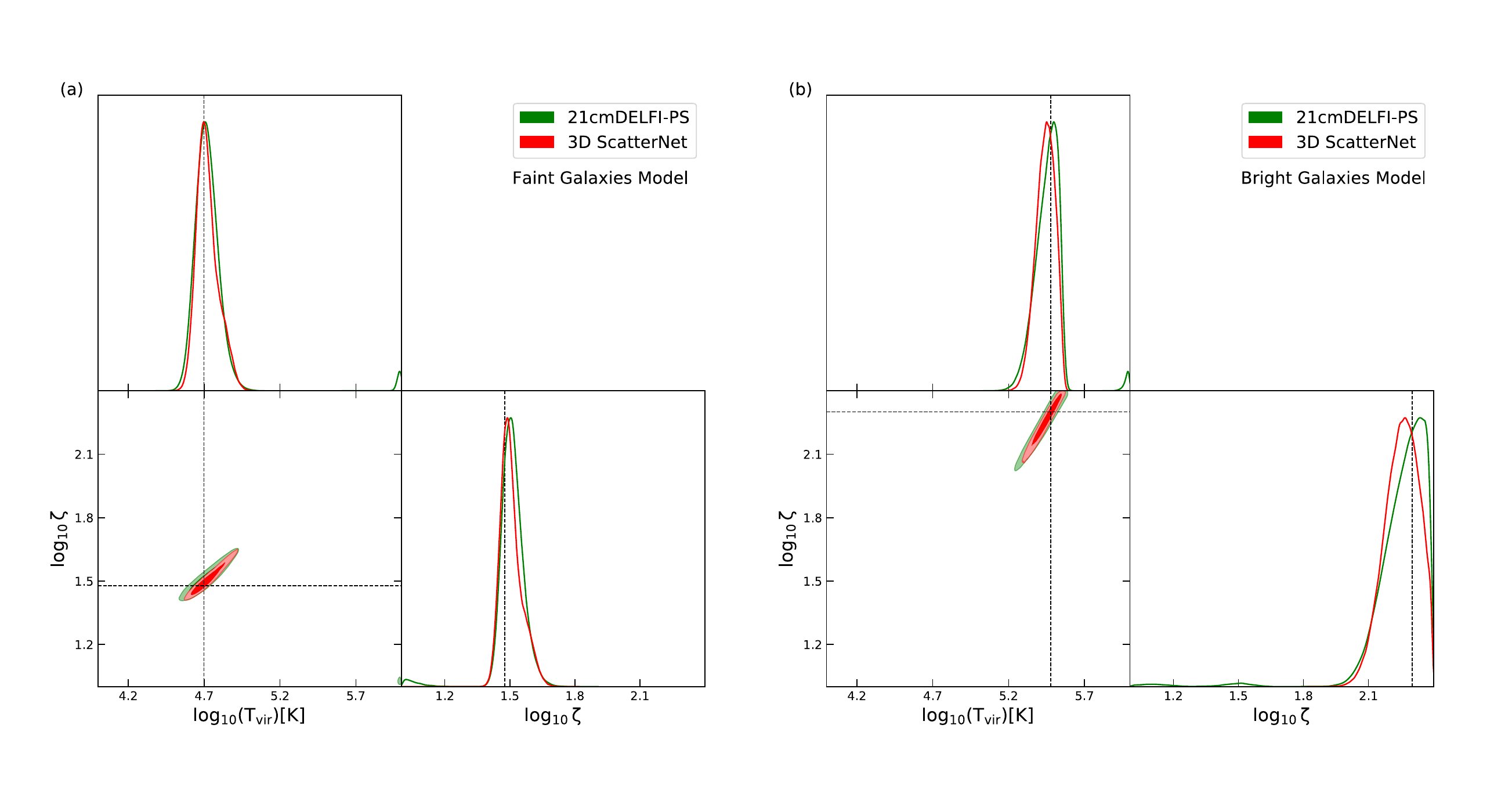}
    \caption{Same as Figure~\ref{fig:comp_pure}, but the estimations are made from mock observations of the 3D 21~cm light-cone images from SKA, which includes the total noise (with the contributions from thermal noise and sample variance errors) yet without foreground contamination. Here we only compare two different approaches of data compression --- power spectrum using {\tt 21cmDELFI-PS} (green) and solid harmonic WST using {\tt 3D ScatterNet} (red). }
    \label{fig:noise_only}
\end{figure*}

\subsubsection{SKA Noise and Residual Foreground}
\label{sec:results_thermal}

We now consider the case of ``SKA noise'' which refers to the mock SKA observations of the 3D 21 cm images with total noise (with the contributions from thermal noise and sample variance errors) yet without foreground contamination. Since the images after smoothing with 1-km baseline lose small-scale information, we discard the components of scattering coefficients with $j=0$ and hence the final scattering coefficients have the dimension of 48. For the power spectrum, we also discard the largest $k$-modes in a single box and generate the final power spectrum vector with a dimension of 120. 

For the case of ``SKA noise'', Figure~\ref{fig:noise_only} shows the credible regions for the ``Faint Galaxies Model'' and the ``Bright Galaxies Model'', with quantitative comparisons in terms of medians and $1\sigma$ errors presented in Tables~\ref{tab: stat_a} and \ref{tab: stat_b}. The inference results of the  ``SKA noise'' case have larger uncertainties than those of ``pure signal'', which is reasonable, given the noise. The statistical errors of {\tt 3D ScatterNet} are smaller than those of {\tt 21cmDELFI-PS}. This improvement is also verified by Table~\ref{tab:perf_comparison}: the $R^2$ values of {\tt 3D ScatterNet} are larger than those of {\tt 21cmDELFI-PS}, and the typical fractional error of the former is about $1.49$ ($1.50$) times smaller for $T_{\rm vir}$ ($\zeta$) than that of the latter. 
 
 Lastly, we consider the case of ``SKA noise + residual foreground'' which refers to the mock SKA observations of the 3D 21~cm images with total noise and residual foreground after the foreground is removed with the SVD technique. For the same reason of smoothing as in the ``SKA noise'' case, we reduce the dimension of the scattering coefficients to be 48 by discarding the components with $j=0$. For the power spectrum, in order to train reliable NDEs, we have to further discard more large-$k$ modes with the upper limit $k_{\rm max}=0.45 \, {\rm Mpc}^{-1}$ and the final vector of power spectrum has the dimension of 70. In this case, the information for parameter inference is only from large-scale modes because the images are smoothed with a rather coarse resolution. 
  
\begin{figure*}
    \centering
    \includegraphics[width=\textwidth]{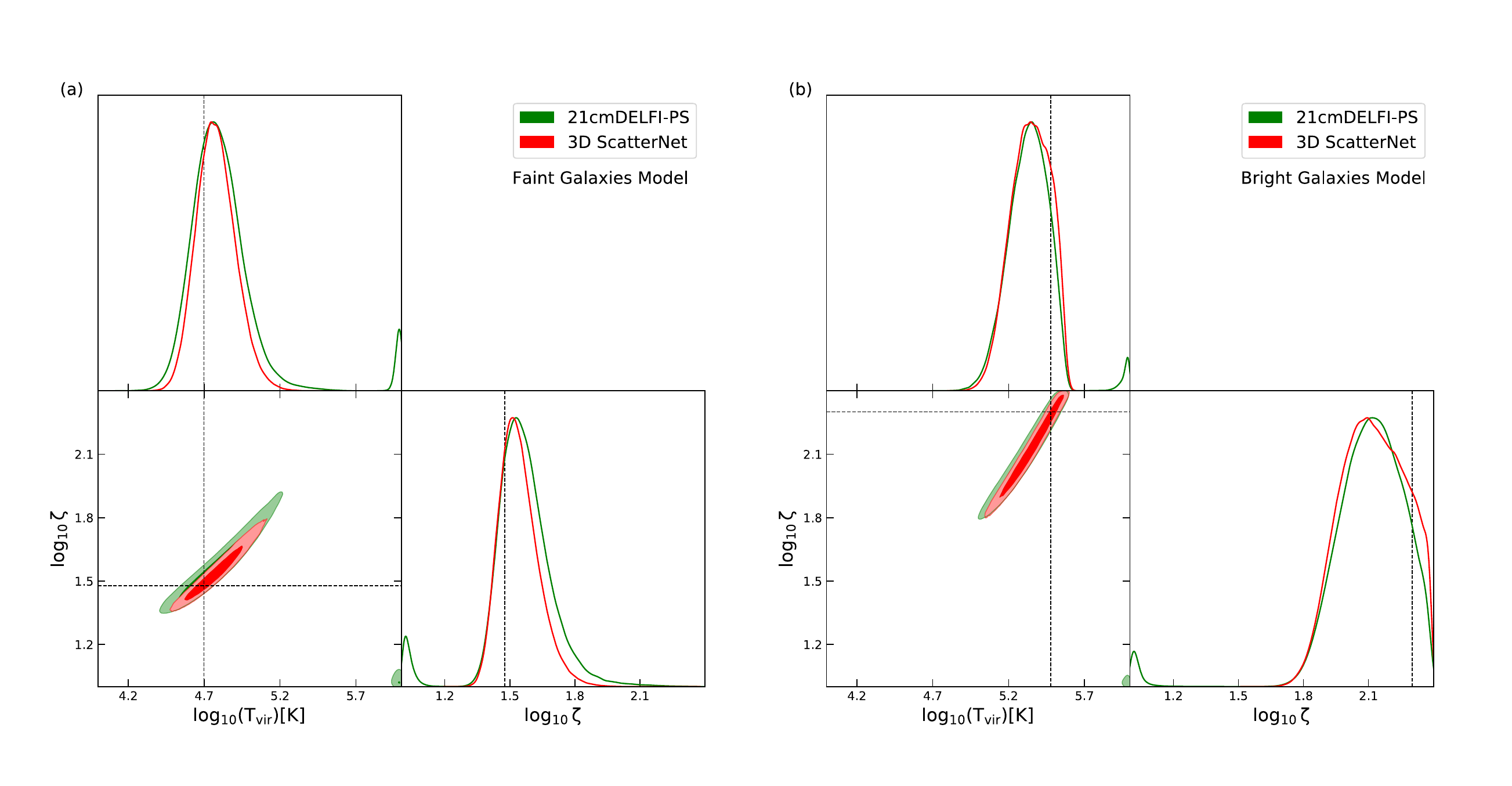}
    \caption{Same as Figure~\ref{fig:noise_only}, but with mock observations of the 3D 21~cm light-cone images from SKA, which includes the total noise and residual foreground after the foreground is removed using SVD.}
    \label{fig:noised}
\end{figure*}

For the case of ``SKA noise + residual foreground'', Figure~\ref{fig:noised} shows the inference results for two mock observations, with the medians and $1\sigma$ errors listed in Tables~\ref{tab: stat_a} and \ref{tab: stat_b}. As expected, given more uncertainties due to residual foreground, the posteriors have larger errors in this case than those of ``SKA noise''. The statistical errors of {\tt 3D ScatterNet} are on average smaller than those of {\tt 21cmDELFI-PS}. This improvement is also verified by Table~\ref{tab:perf_comparison}: the $R^2$ values of {\tt 3D ScatterNet} are higher than those of {\tt 21cmDELFI-PS}, and the typical fractional error of the former is about $1.09$ ($1.15$) times smaller for $T_{\rm vir}$ ($\zeta$) than that of the latter. 
 
 Regarding the posterior validation, all $p$-values are larger than 0.01 both for the ``SKA noise'' case and the case of ``SKA noise + residual foreground'', which implies that our results are at least reliable with a significance of 0.01. Also, the $p$-values of {\tt 3D ScatterNet} are generically larger than those of {\tt 21cmDELFI-PS}. 
 
In sum, for all cases of different assumptions in noise and residual foreground, {\tt 3D ScatterNet} outperforms the {\tt 21cmDELFI-PS}. Our results demonstrate that the solid harmonic WST can extract the information from the 3D 21~cm light-cone images in a more effective way than the power spectrum statistic itself. These results are consistent with the previous findings that the $n$th-order scattering transform captures the information up to the $2^n$-point function \citep{mallat2012group,2021arXiv211201288C}. In addition, \citet{2023arXiv230704994S} introduced the concept of \emph{mutual information} to evaluate the comparisons between different data summaries by giving them scores and found that the solid harmonic WST not only outperforms the power spectrum but also surpasses the combined information including both power spectrum and bispectrum, in the context of extracting the information from the 21~cm datacube.

\subsection{Discussions} 
\label{sec:discuss_noise}

In this subsection, we attempt to provide some insights regarding why the 3D CNN (trained with the particular set of experimental choices in \citetalias{2021arXiv210503344Z}) performs less effectively than the solid harmonic WST (and power spectrum) in compressing the 3D 21~cm light-cone images, and the limitations of the 3D CNN configurations in \citetalias{2021arXiv210503344Z}.

The solid harmonic WST is analogous to a 3D CNN, but unlike the latter, it essentially gives a fixed kernel without the training process. 
Furthermore, the resulting scattering coefficients are naturally invariant to translations and rotations and are Lipschitz continuous to deformations \citep{https://doi.org/10.1002/cpa.21413}. These properties are particularly useful for 3D 21~cm light-cone images because the data summaries have the translational invariance in the angular direction and the rotational invariance along the LOS\footnote{In practice, we obtain approximate invariance that is limited by the boundaries and the discrete sampling of images.}, and can be stable under slight deformations due to variations in the initial density fields. These intrinsic invariance properties can enhance conditional density learning of the solid harmonic WST.

In principle, thanks to the similarity between solid harmonic WST and 3D CNN, a deep and/or wide 3D CNN should perform at least as well as the WST in extracting information, as long as three conditions are all met --- training data is adequate, network is sufficiently expressive, and training process is successful. Our results imply that the 3D CNN configurations in \citetalias{2021arXiv210503344Z} might have limitations in these three aspects.

The first limitation is insufficient training data. A dataset that samples the complete parameter-data distribution is required to train a robust neural network. For training a 2D CNN \citep{Gillet2019}, a dataset comprising 9,000 training samples might suffice, but the same number of training samples is likely not adequate for 3D CNN\footnote{We note that \citet{2022MNRAS.511.3446N} trained a 3D CNN with a smaller number of data than \citetalias{2021arXiv210503344Z}, but it performed comparably well. This might benefit from the specific architecture adopted therein.}. While we find that the results for the ``Faint Galaxies Model'' and the ``Bright Galaxies Model'' were not improved if the training sample size was doubled, a thorough scaling test on the number of training samples is necessary to make a robust diagnosis. 

The second limitation is the insufficient complexity of neural network architectures. 
While the solid harmonic WST exploits its intrinsic invariance properties, 
the standard CNNs cannot because they are not invariant to rotations due to inherent architecture. 
Thus the architecture in CNNs needs to be fine-tuned during the training. Alternatively, a likely more effective approach is to integrate additional inductive biases into the architecture \citep{2018arXiv180101450K,9156444,NEURIPS2018_488e4104,8578193}. In these variants of CNNs, the invariance properties can be enhanced, which might optimize the network architecture. Also, in \citet{2022MNRAS.511.3446N}, an elongated kernel is designed to detect the patterns evolving in the redshift direction, which reduces the number of trainable parameters\footnote{Similarly in this work, we also use wavelet filters with elongated shapes for solid harmonic WST. \citet{2022MNRAS.509.3852P} included some discussions on the neural network architectures used for interpreting the 21~cm lightcones.}.

The third limitation is the sub-optimal network training process. Network optimization involves refined practices for neural network initialization and learning rate determination \citep{2023arXiv230611668D}. Some generic optimization strategies \citep{2022MNRAS.511.3446N,2019arXiv190710902A} may be useful as well.

Note that the 3D CNN trained in \citetalias{2021arXiv210503344Z} performs even less effectively than the power spectrum analysis using {\tt 21cmDELFI-PS}. This might be explained by the limitations of 3D CNN discussed above. While the 3D CNN configurations in \citetalias{2021arXiv210503344Z} might be improved in these three aspects, we leave a thorough exploration along these lines to future work.

The comparison involving 3D CNNs in this work has been limited to the case of pure signal. This is because the comparison when instrumental effects are applied would require additional fine-tuning efforts for the 3D CNN to fully assess its potential, which is not the focus of this work. Appendix~\ref{sec:appendix_noise} provides some insight into such a comparison by including the SKA noise in the application of {\tt DELFI-3DCNN}, nevertheless without additional fine-tuning, i.e.\ using the same hyper-parameters for training the 3D CNN as detailed in \citetalias{2021arXiv210503344Z}.

\section{Summary}
\label{sec:conclusion}

In this paper, we introduce the solid harmonic WST for compressing the 3D image data and generating meaningful low-dimensional summaries. We apply this technique to the data compression of 3D tomographic 21~cm light-cone images and use the resulting scattering coefficients as the input summaries of the DELFI. With DELFI, we perform the Bayesian inference of the reionization parameters where the likelihood is implicitly defined by the forward simulations. This new technique, dubbed {\tt 3D ScatterNet} (i.e.\ solid harmonic WST with DELFI), recovers accurate posterior distributions for the
reionization parameters.  

We compare the inference results with two different approaches of data compression, the solid harmonic WST and the 3D CNN, for two representative mock observations in the ``pure signal'' case, and demonstrate that the {\tt 3D ScatterNet} outperforms the {\tt DELFI-3D CNN} significantly. Our results imply that the solid harmonic WST extracts the information from the 3D 21~cm light-cone images in a more informative manner than a 3D CNN given reasonable fine-tuning. We also offer insights into enhancing the 3D CNN regarding the datasets, network architecture, and training process. Moreover, the solid harmonic WST has fixed kernels in wavelets, which means that it does not need to be trained in order to output the data summaries. This highlights its robustness and efficiency, which is another advantage of the solid harmonic WST over the 3D CNN.

We then make another comparison of the inference results between data compression using the solid harmonic WST ({\tt 3D ScatterNet}) and using the power spectrum statistic ({\tt 21cmDELFI-PS}), both under the same inference strategy using DELFI. The comparisons were made for three cases with different assumptions on the noise and residual foreground --- ``pure signal'', ``SKA noise'' and ``SKA noise + residual foreground''. We find that the {\tt 3D ScatterNet} outperforms the {\tt 21cmDELFI-PS} in all cases. This implies that the summaries compressed with the solid harmonic WST contain more (i.e.\ non-Gaussian) information from the 3D 21~cm light-cone images than the power spectrum analysis. 

Our results demonstrate that combining a WST (3D solid harmonic WST in this paper) with the simulation-based inference will be a promising tool for the scientific interpretation of future 21~cm light-cone image observation data. Based on the findings of this paper, there is room for improvement with regard to the design of summary statistics from solid harmonic WST or new WST. For example, instead of treating the light-cone as a whole to build a statistic, a variant approach might be to treat discrete boxes separately and concatenate the scattering coefficients of each box together as the data summaries. In this way, the correlations between different stages of reionization can be exploited by the CMAFs. Nevertheless, this will increase the dimension of data summaries and therefore incur higher computational costs, so the trade-off between accuracy and efficiency in the solid harmonic WST is yet to be further explored. Also, further cross-correlation between scales resulting from re-scaling the phase information is likely to be exploited in a 3D extension of the wavelet phase harmonics (WPH; \citealt{10.1093/imaiai/iaz019,allys2020new}) for the 2D case. On the other hand, when scaling the similar analysis to a high-dimensional parameter and/or data space, the diffusion model \citep{2022arXiv220811970L, 2023arXiv230403788L, 2023MNRAS.526.1699Z} is a promising alternative to the CMAFs, or other normalizing flows, to perform the simulation-based inference because of its simplified training objective. We leave the exploration of these directions to future works.

{\it Note. -- } In the final stage of preparing this manuscript, \citet{2022arXiv220402544G,2022arXiv220407646E} were posted on arXiv. These papers applied the scattering transform (or wavelet moments, varied band-limited first-order scattering coefficients defined in the Fourier space, in \citealt{2022arXiv220407646E}) for parameter inference, but both used the Fisher matrix formalism. In comparison, we apply the scattering transform for parameter inference using simulation-based inference in this paper. In addition, \citet{2022arXiv220402544G} focuses on the reionization parameter estimation using the 2D 21~cm images; \citet{2022arXiv220407646E} focuses on the cosmological parameter estimation using the 3D cosmological density fields. In comparison, our paper has a distinct focus, i.e.\ on the reionization parameter estimation using the 3D 21~cm light-cone images.

\section*{Acknowledgements}
This work is supported by the National SKA Program of China (grant No.~2020SKA0110401), NSFC (grant No.~11821303), and the National Key R\&D Program of China (grant No.~2018YFA0404502). BDW acknowledges support from the Simons Foundation. 
We thank Sihao Cheng, Paulo Montero-Camacho, and Bohua Li for their useful discussions and help. We acknowledge the Tsinghua Astrophysics High-Performance Computing platform at Tsinghua University for providing computational and data storage resources that have contributed to the research results reported within this paper. 

\software{21CMMC \citep{2015MNRAS.449.4246G,2017MNRAS.472.2651G,Greig2018}, 21cmFAST \citep{Mesinger2007,Mesinger2011}, pydelfi \citep{alsing2019fast}, TensorFlow \citep{abadi2016tensorflow}, GetDist \citep{Lewis:2019xzd}, NumPy \citep{harris2020array}, Matplotlib \citep{Hunter:2007}, SciPy \citep{2020SciPy-NMeth}, scikit-learn \citep{Scikit-learn}, Python2 \citep{van1995python}, Python3 \citep{py3}, Kymatio \citep{2018arXiv181211214A}, Tools21cm \citep{Giri2020}, pygsm \citep{2008MNRAS.388..247D,zheng2017improved}, Mayavi \citep{ramachandran2011mayavi}, galpro \citep{mucesh2021machine}, seaborn \citep{Waskom2021}, Astropy \citep{2013A&A...558A..33A, 2018AJ....156..123A}, healpy \citep{Zonca2019}, HEALPix \citep{2005ApJ...622..759G}.}

\appendix

\begin{figure*}
    \centering
    \includegraphics[width=\textwidth]{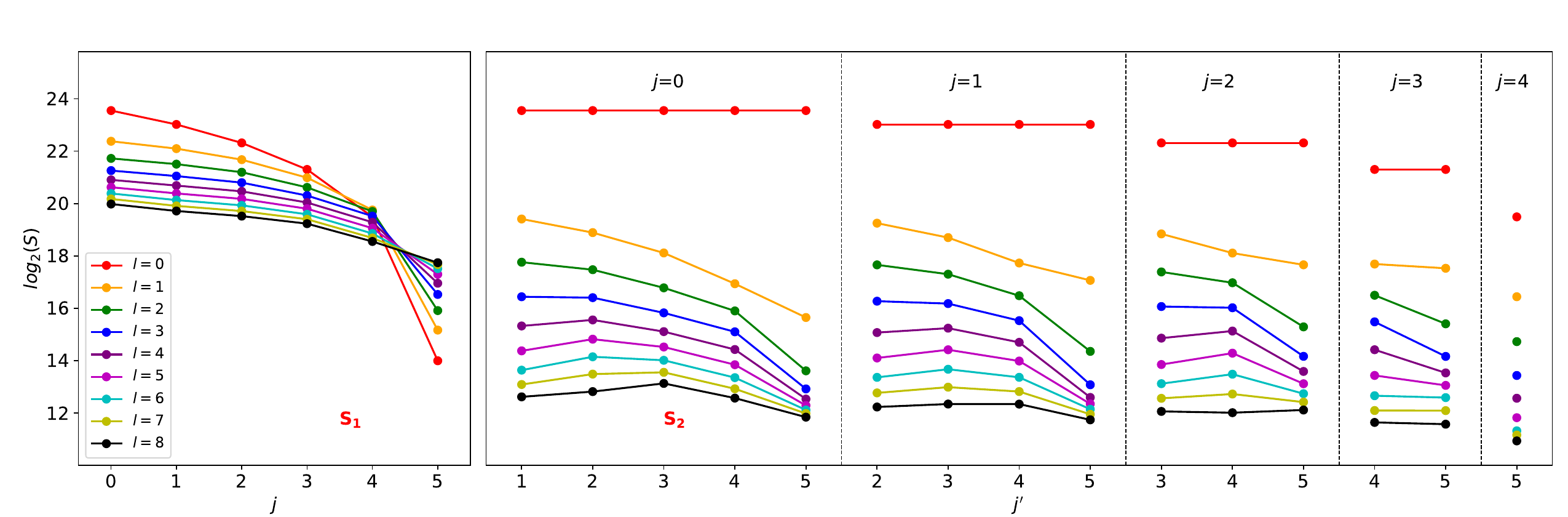}
    \caption{The first-order (``$\mathbf{S_1}$'') and second-order (``$\mathbf{S_2}$'') scattering coefficients of the simulated cosmological 21~cm signals, evaluated at given values of $\ell$ and at $q=1$.}
    \label{fig:statistics_l}
\end{figure*}

\begin{figure}
    \centering
    \includegraphics[width=0.5\textwidth]{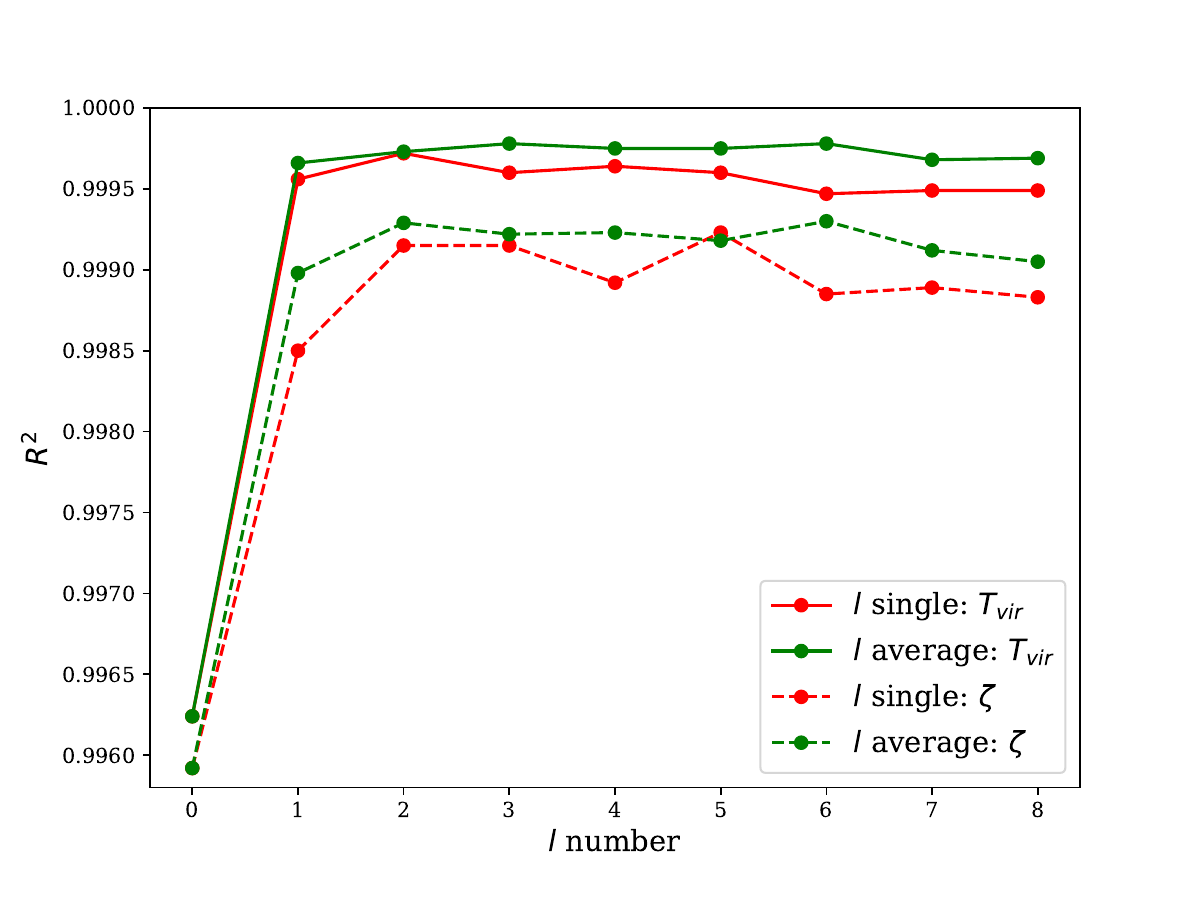}
    \caption{The coefficient of determination $R^2$ as a function of the angular frequency $\ell$. The phrase ``$\ell$ single'' refers to the scenario where the scattering coefficients are evaluated at a given single value of $\ell$, while the phrase  ``$\ell$ average'' refers to the scenario where the scattering coefficients are averaged over $0\le \ell ' \le \ell$ for a given value of $\ell$. The parameter after this phrase (i.e.\ $T_{\rm vir}$ or $\zeta$) indicates the parameter of the $R^2$ calculation.}
    \label{fig:results_l}
\end{figure}

\section{The effect of angular frequency $\ell$ }
\label{sec:l_number}
In the main text of this paper, the scattering coefficients are averaged over $0 \le \ell \le \ell_{\rm max}$ with $\ell_{\rm max}=6$. In this section, we explore the effect of the angular frequency $\ell$ on the parameter inference. 

Figure~\ref{fig:statistics_l} shows the scattering coefficients evaluated at given values of $\ell$. While the first-order scattering coefficients show tilting when varying $\ell$, the second-order scattering coefficients decrease in overall amplitude as $\ell$ increases. Also, for $\ell=0$ (i.e.\ Gaussian wavelets), the scattering coefficients are flat at the scale of $j^\prime$. 

We construct two sets of experiments of posterior inference over 300 testing samples and show the coefficient of determination $R^2$ of the predicted medians in Figure~\ref{fig:results_l}. In the first set of experiments (``$\ell$ single''), the scattering coefficients are evaluated at each single value of $\ell$. We find that using the $\ell=0$ information has the smallest values of $R^2$. When using the information at larger $\ell$, the $R^2$ value increases and then slightly decreases. This implies that the solid harmonic wavelet has better performance than the Gaussian wavelet (i.e.\ the case with $\ell =0$).  In the second set of experiments (``$\ell$ average''), the scattering coefficients are averaged over $0\le \ell ' \le \ell$ for a given value of $\ell$. Comparing these two scenarios, we find that the combined information (``$\ell$ average'') gets a higher value of $R^2$ generically than the single information (``$\ell$ single''). When increasing the value of $\ell$, in the case of ``$\ell$ average'', the $R^2$ value increases and then slightly decreases, with the peak at $\ell=6$. For this reason, in the main text of this paper, we choose the ``$\ell$ average'' scenario with $\ell_{\rm max}=6$.

\section{The effect of smoothing scales}
\label{sec:comp_km}

In the main text of this paper, the images are smoothed with the scale corresponding to the size of the 1 km baseline. In this section, we test the effect of smoothing scales on parameter inference. We compute the values of $R^2$ and typical fractional error $\epsilon$ for the predicted medians of 300 testing samples. Table~\ref{tab:smooth} shows that smoothing with the 1-km baseline has a larger $R^2$ value and smaller fractional error than smoothing with the 2-km baseline. This is because the smoothed images in the former have higher S/N although in coarser resolution. This result is the reason why we chose the size of the 1-km baseline for smoothing in this paper. 

\begin{table}
    \centering
    \caption{Recovery Performance by the {\tt 3D ScatterNet} for Smoothing with the Baseline of 1-km and 2-km Respectively}
    \begin{tabular}{cccccc} 
    \hline \hline
	& &\multicolumn{2}{c}{\scriptsize{SKA noise}} &\multicolumn{2}{c}{\scriptsize{SKA noise + residual foreground}} \\
			 \cmidrule(l{.75em}r{.75em}){3-4}
			 \cmidrule(l{.75em}r{.75em}){5-6}
	& &\multicolumn{1}{c}{\scriptsize{1 km}} &\multicolumn{1}{c}{\scriptsize{2 km}} &\multicolumn{1}{c}{\scriptsize{1 km}}&\multicolumn{1}{c}{\scriptsize{2 km}}\\
        \hline
\multirow{2}{*}{\scriptsize{$R^2$}} & \scriptsize{$ \log_{10}(T_{\rm vir})$} & \scriptsize{0.9647} & \scriptsize{0.8808} & \scriptsize{0.8348} & \scriptsize{0.7853}  \\ 
{} & \scriptsize{$\log_{10}(\zeta)$} & \scriptsize{0.9604} & \scriptsize{0.9153} & \scriptsize{0.8254} & \scriptsize{0.8212} \\
    \hline
 \multirow{2}{*}{\scriptsize{$\epsilon$} } & \scriptsize{$ T_{\rm vir}$} & \scriptsize{$(-0.17, 0.16)$} & \scriptsize{$(-0.18, 0.27)$} & \scriptsize{$(-0.28, 0.41)$} & \scriptsize{$(-0.27, 0.44)$} \\ 
{} & \scriptsize{$\zeta$} & \scriptsize{$(-0.12, 0.14)$} & \scriptsize{$(-0.16, 0.20)$} & \scriptsize{$(-0.22, 0.32)$} & \scriptsize{$(-0.25, 0.29)$} \\
\hline 
    \end{tabular}
  \label{tab:smooth}
\end{table}

\begin{figure*}
    \centering
    \includegraphics[width=\textwidth]{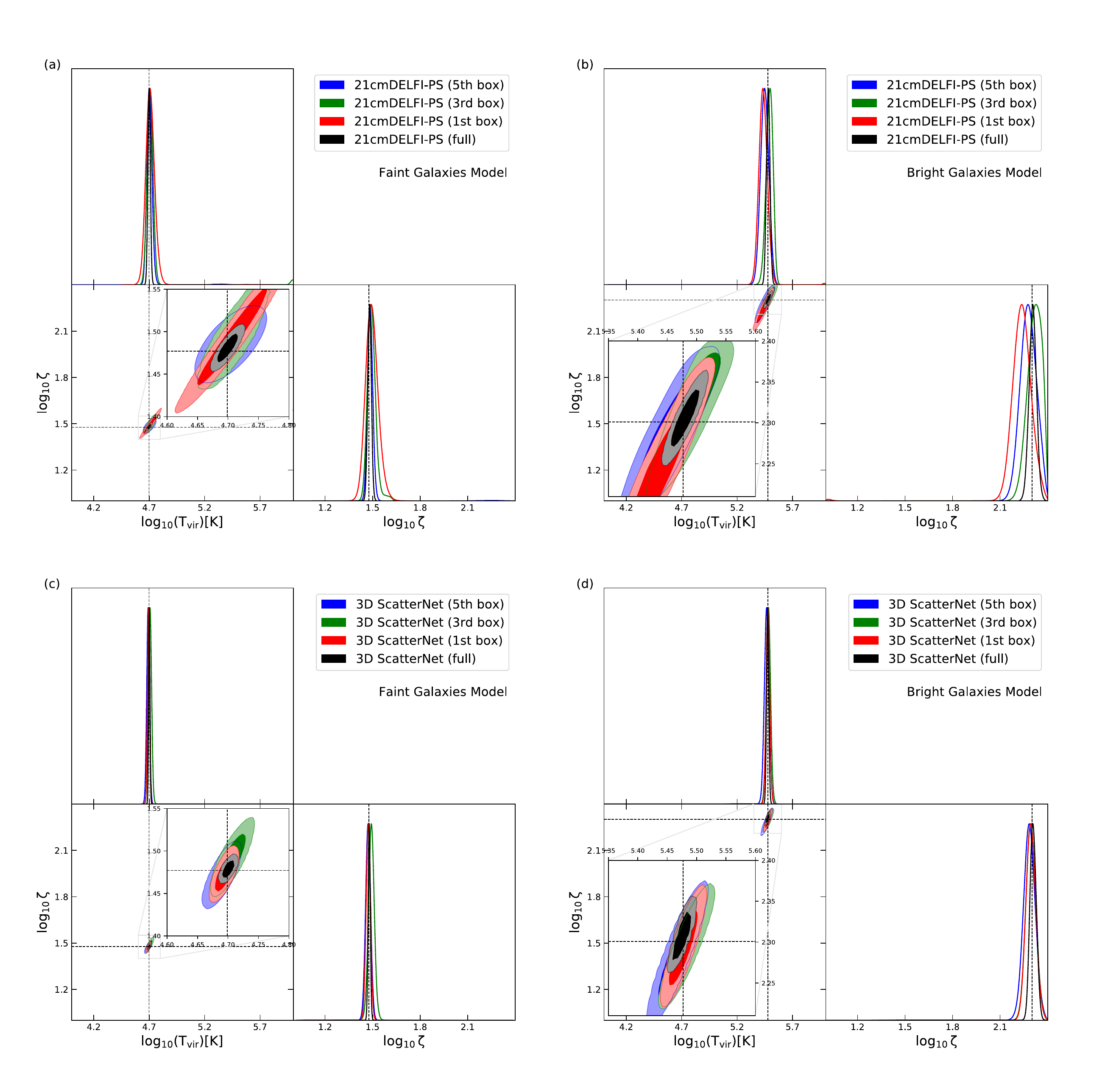}
    \caption{The posteriors estimated from 3D cosmic 21~cm light-cone images (i.e.\ ``pure signal'') for the scenario where only the information within the first (red), third (green), fifth (blue) box and the full (black) light-cone images is exploited for parameter inference, respectively. Shown are the results using {\tt 21cmDELFI-PS} (top) and using {\tt 3D ScatterNet} (bottom) for the ``Faint Galaxies Model'' (left) and the ``Bright Galaxies Model'' (right), respectively.}
    \label{fig:single}
\end{figure*}

\begin{table*}
    \centering
    \caption{Recovery Performance for Information in Discrete Boxes}
    \begin{tabular}{cccccccccc}
    \hline \hline
	& &\multicolumn{4}{c}{\scriptsize{{\tt 21cmDELFI-PS}}} &\multicolumn{4}{c}{\scriptsize{{\tt 3D ScatterNet}}} \\
			 \cmidrule(l{.75em}r{.75em}){3-6}
			 \cmidrule(l{.75em}r{.75em}){7-10}
			  & &  \scriptsize{First} & \scriptsize{Third} & \scriptsize{Fifth} & \scriptsize{Full} &   \scriptsize{First} & \scriptsize{Third} & \scriptsize{Fifth} & \scriptsize{Full} \\
        \hline
\multirow{2}{*}{\scriptsize{$R^2$}} & \scriptsize{$ \log_{10}(T_{\rm vir})$} & \scriptsize{*$^\dag$} & \scriptsize{0.9948} & \scriptsize{0.9863} & \scriptsize{0.9989} & \scriptsize{0.9991} & \scriptsize{0.9987} & \scriptsize{0.9978} & \scriptsize{0.9997} \\ 
{} & \scriptsize{$\log_{10}(\zeta)$} & \scriptsize{*} & \scriptsize{0.9909} & \scriptsize{0.9754} & \scriptsize{0.9978} & \scriptsize{0.9960} & \scriptsize{0.9953} & \scriptsize{0.9924} & \scriptsize{0.9990} \\
    \hline
 \multirow{2}{*}{\scriptsize{$\epsilon$} } & \scriptsize{$ T_{\rm vir}$} & \scriptsize{*} & \scriptsize{$(-0.06, 0.08)$} & \scriptsize{$(-0.07, 0.08)$} & \scriptsize{$(-0.03, 0.03)$} & \scriptsize{$(-0.03, 0.03)$} & \scriptsize{$(-0.03, 0.04)$} & \scriptsize{$(-0.05, 0.04)$} & \scriptsize{$(-0.02, 0.02)$}  \\ 
{} & \scriptsize{$\zeta$} & \scriptsize{*} & \scriptsize{$(-0.05, 0.08)$} & \scriptsize{$(-0.08, 0.07)$} & \scriptsize{$(-0.04, 0.03)$} & \scriptsize{$(-0.04, 0.05)$} & \scriptsize{$(-0.04, 0.05)$} & \scriptsize{$(-0.06, 0.04)$} & \scriptsize{$(-0.02, 0.02)$} \\
\hline 
    \end{tabular}
  \label{tab:single}
		\flushleft
		\tablenotetext{}{{\bf Note.} The parameter inference is made for the case of ``pure signal''. Here, ``First'', ``Third'' and ``Fifth'' refer to the scenario where only the information within the first, third, and fifth box (counting from the low-redshift, corresponding to the central redshift of $7.67$, $8.36$, and $9.11$) is exploited for parameter inference, respectively. ``Full'' refers to the case of exploiting the full light-cone images (i.e.\  concatenating all ten boxes), which is the case in the main text of this paper. $^\dag$ We do not show these results for the first box because the hypothesis test with the HPD values failed, likely due to the lack of training samples. }
\end{table*}

\section{The dependence of light-cone volumes}
\label{sec:single}

In the main text of this paper, the full light-cone images, which concatenate ten light-cone boxes, are exploited for parameter inference. In this section, we investigate whether the information from one of these boxes dominates over the others. (If this were the case, then observations could be optimized targeting at the ``sweet-spot'' redshift.) 

For this purpose, we perform the parameter inference from discrete boxes, both using the power spectrum and using scattering coefficients as summaries. Figure~\ref{fig:single} shows the posteriors for two mock observations and Table~\ref{tab:single} shows the results of recovery performance for the predicted medians of 300 testing samples. For the {\tt 3D ScatterNet}, the box at the low-redshift results in better inference performance in terms of larger $R^2$, smaller typical fractional error, and smaller credible region for mock observations than that at the high-redshift. However, no single box appears to dominate the information, because the full light-cone images always yield the optimal result that is significantly better than any of the single boxes. These results hold similarly for the {\tt 21cmDELFI-PS}. We conclude that there is no optimal redshift. As such, the full light-cone images, which gather the complete information, provide the parameter inference with the best performance. 

\section{The NDE setting and sample size}
\label{sec:nde_setting}

In this section, we give the details of the networks as follows. We choose the CMAFs as the NDEs throughout of this paper. In all architectures, we set two neural layers of a single transform, represented by the masked autoencoders for density estimation (MADE; \citealt{germain2015made}), with 50 neurons per layer. We also use the ensembles of NDEs. The final posterior is the stacked one from individual posteriors with weights according to their training errors. We fine-tune the training sample size and the CMAFs architecture for every method and experiment based on the outcomes of posterior validation (calibration). The guiding principle behind this fine-tuning is that data (summaries) with greater uncertainties might require more intricate CMAFs, such as increased transformations within a single CMAF and the increased number of CMAFs within the ensemble. Meanwhile, data (summaries) of higher dimensions might necessitate a more substantial training sample due to the vast feature space.

The procedure for scattering coefficients is as follows. 
\begin{itemize}
	\item [(1)] The scattering coefficients for the ``pure signal'' case with ${\bf k}_\perp = 0$ mode removed (see Section~\ref{sec:results_pure}). We use 18,000 light-cone images and an ensemble of 8 NDEs $(5,6,7,8)*2$, which means that we have two NDE blocks, each having the NDE with the number of transformations 5, 6, 7, and 8, respectively. Hereafter we use the same terminology. Notice that the final performance can be further enhanced with double the size of the light-cone images. To fairly compare with {\tt 21cmDELFI-PS}, we use 18,000 images for reports in the paper.
	\item [(2)] The coefficients from pure discrete boxes (see Appendix~\ref{sec:single}). We use 27,000 samples. The ensembles for the first, third, and fifth boxes: $(5,6,7,8)*3$, $(5,6,7,8)*3$, $(5,6,7,8)*2$, respectively.
	\item [(3)] The coefficients with the information of averaged $\ell$ from pure boxes (see Appendix~\ref{sec:l_number}). We use 36,000 samples. The ensembles from $\ell=1$ to $\ell=8$: $(5)*4$, $(6,7,8,9)*3$, $(6,7,8,9)*2$, $(5,6,7,8)*2$, $(6,7,8,9)*3$, $(5,6,7,8)*2$, $(5)*4$, $(5,6,7,8)*2$, respectively.
	\item [(4)] The coefficients with the information of single $\ell$ from pure boxes (see Appendix~\ref{sec:l_number}). We use 36,000 samples. The ensembles from $\ell=0$ to $\ell=8$: $(5)*4$, $(5)*4$, $(5,6,7,8)*2$, $(5)*4$, $(5,6,7,8)*2$, $(5,6,7,8)*2$, $(5,6,7,8)*2$, $(5,6,7,8)*2$, $(5,6,7,8)*2$, respectively.
	\item [(5)] The coefficients from the signal with SKA noise. We use 27,000 samples (see Section~\ref{sec:results_thermal}). The ensembles for the signal with 1-km smoothing: $(5,6,7,8)*2$, and 2-km smoothing: $(5,6,7,8)*3$, respectively.
	\item [(6)] The coefficients from the signal with SKA noise and residual foreground (see Section~\ref{sec:results_thermal}). We use 36,000 samples. The ensembles for the signal with 1-km smoothing: $(20)*2$, and 2-km smoothing: $(30)*2$, respectively.
\end{itemize} 

The procedure for the power spectrum is as follows. 
\begin{itemize}
	\item [(1)] The power spectrum from pure signal with ${\bf k}_\perp = 0$ mode removed (see Section~\ref{sec:results_pure}). We use 18,000 samples and the ensembles $(5)*4$.
	\item [(2)] The power spectrum from pure discrete boxes (see Appendix~\ref{sec:single}). For the first box, we use 32,596 samples and the ensembles $(6,7,8,9)*3$; for the third box, we use 34,044 samples and the ensembles $(5,6,7,8)*3$; and for the fifth box, we use 27,000 samples and the ensembles $(5,6,7,8)*2$.
	\item [(3)] The power spectrum from the signal with SKA noise (see Section~\ref{sec:results_thermal}). We use 27,000 samples and the ensembles $(5)*4$.
	\item [(4)] The power spectrum from the signal with SKA noise and residual foreground (see Section~\ref{sec:results_thermal}). We use 36,000 samples and the ensembles $(20)*3$.
\end{itemize}

\section{Performance of DELFI-3DCNN with SKA noise}
\label{sec:appendix_noise}

\begin{figure*}
    \centering
    \includegraphics[width=\textwidth]{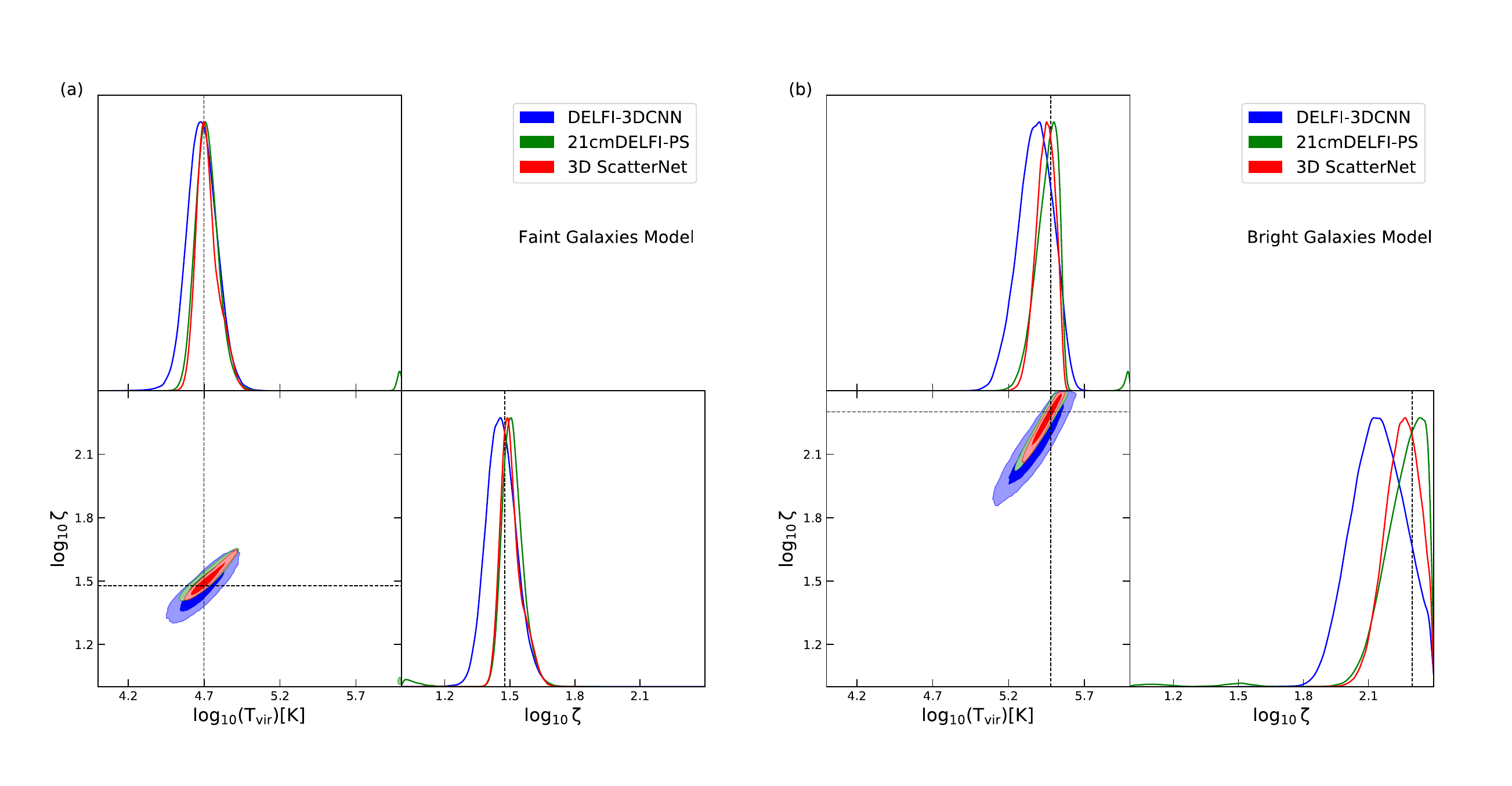}
    \caption{Same as Figure~\ref{fig:noise_only} but including the results from {\tt DELFI-3DCNN}.}
    \label{fig:appendix_noise}
\end{figure*}

In this section, we extend the comparison of 3D CNN using {\tt DELFI-3D CNN}, power spectrum using {\tt 21cmDELFI-PS} and solid harmonic WST using {\tt 3D ScatterNet} to the mock observations of the 3D 21~cm light-cone images from SKA, which includes
the total noise yet without foreground contamination. Ideally, this would require additional fine-tuning efforts for the 3D CNN to fully assess its potential. Here we present the results of {\tt DELFI-3D CNN} without additional fine-tuning, i.e.\ using the same hyper-parameters for training the 3D CNN and NDEs as detailed in \citetalias{2021arXiv210503344Z}. Figure~\ref{fig:appendix_noise} shows that in this realistic scenario, a trained 3D CNN still performs sub-optimal compared to {\tt 21cmDELFI-PS} and {\tt 3D ScatterNet}.

\bibliographystyle{aasjournal}
\bibliography{Ref}{}

\begin{thebibliography}{}
\expandafter\ifx\csname natexlab\endcsname\relax\def\natexlab#1{#1}\fi
\providecommand{\url}[1]{\href{#1}{#1}}
\providecommand{\dodoi}[1]{doi:~\href{http://doi.org/#1}{\nolinkurl{#1}}}
\providecommand{\doeprint}[1]{\href{http://ascl.net/#1}{\nolinkurl{http://ascl.net/#1}}}
\providecommand{\doarXiv}[1]{\href{https://arxiv.org/abs/#1}{\nolinkurl{https://arxiv.org/abs/#1}}}

\bibitem[{Abadi {et~al.}(2016)Abadi, Barham, Chen, Chen, Davis, Dean, Devin, Ghemawat, Irving, Isard, Kudlur, Levenberg, Monga, Moore, Murray, Steiner, Tucker, Vasudevan, Warden, Wicke, Yu, \& Zheng}]{abadi2016tensorflow}
Abadi, M., Barham, P., Chen, J., {et~al.} 2016, in Proceedings of the 12th USENIX Conference on Operating Systems Design and Implementation, OSDI'16 (Berkeley, CA: USENIX Association), 265–283

\bibitem[{{Abdurashidova} {et~al.}(2022){Abdurashidova}, {Aguirre}, {Alexander}, {Ali}, {Balfour}, {Beardsley}, {Bernardi}, {Billings}, {Bowman}, {Bradley}, {Bull}, {Burba}, {Carey}, {Carilli}, {Cheng}, {DeBoer}, {Dexter}, {de Lera Acedo}, {Dibblee-Barkman}, {Dillon}, {Ely}, {Ewall-Wice}, {Fagnoni}, {Fritz}, {Furlanetto}, {Gale-Sides}, {Glendenning}, {Gorthi}, {Greig}, {Grobbelaar}, {Halday}, {Hazelton}, {Hewitt}, {Hickish}, {Jacobs}, {Julius}, {Kern}, {Kerrigan}, {Kittiwisit}, {Kohn}, {Kolopanis}, {Lanman}, {La Plante}, {Lekalake}, {Lewis}, {Liu}, {MacMahon}, {Malan}, {Malgas}, {Maree}, {Martinot}, {Matsetela}, {Mesinger}, {Molewa}, {Morales}, {Mosiane}, {Murray}, {Neben}, {Nikolic}, {Nunhokee}, {Parsons}, {Patra}, {Pascua}, {Pieterse}, {Pober}, {Razavi-Ghods}, {Ringuette}, {Robnett}, {Rosie}, {Sims}, {Singh}, {Smith}, {Syce}, {Thyagarajan}, {Williams}, {Zheng}, \& {HERA Collaboration}}]{2022ApJ...925..221A}
{Abdurashidova}, Z., {Aguirre}, J.~E., {Alexander}, P., {et~al.} 2022, \apj, 925, 221, \dodoi{10.3847/1538-4357/ac1c78}

\bibitem[{{Akiba} {et~al.}(2019){Akiba}, {Sano}, {Yanase}, {Ohta}, \& {Koyama}}]{2019arXiv190710902A}
{Akiba}, T., {Sano}, S., {Yanase}, T., {Ohta}, T., \& {Koyama}, M. 2019, arXiv e-prints, arXiv:1907.10902, \dodoi{10.48550/arXiv.1907.10902}

\bibitem[{{Allys} {et~al.}(2019){Allys}, {Levrier}, {Zhang}, {Colling}, {Regaldo-Saint Blancard}, {Boulanger}, {Hennebelle}, \& {Mallat}}]{allys2019rwst}
{Allys}, E., {Levrier}, F., {Zhang}, S., {et~al.} 2019, \aap, 629, A115, \dodoi{10.1051/0004-6361/201834975}

\bibitem[{{Allys} {et~al.}(2020){Allys}, {Marchand}, {Cardoso}, {Villaescusa-Navarro}, {Ho}, \& {Mallat}}]{allys2020new}
{Allys}, E., {Marchand}, T., {Cardoso}, J.~F., {et~al.} 2020, \prd, 102, 103506, \dodoi{10.1103/PhysRevD.102.103506}

\bibitem[{{Alsing} {et~al.}(2019){Alsing}, {Charnock}, {Feeney}, \& {Wandelt}}]{alsing2019fast}
{Alsing}, J., {Charnock}, T., {Feeney}, S., \& {Wandelt}, B. 2019, \mnras, 488, 4440, \dodoi{10.1093/mnras/stz1960}

\bibitem[{{Alsing} {et~al.}(2018){Alsing}, {Wandelt}, \& {Feeney}}]{alsing2018massive}
{Alsing}, J., {Wandelt}, B., \& {Feeney}, S. 2018, \mnras, 477, 2874, \dodoi{10.1093/mnras/sty819}

\bibitem[{Anderson(1962)}]{10.1214/aoms/1177704477}
Anderson, T.~W. 1962, The Annals of Mathematical Statistics, 33, 1148 , \dodoi{10.1214/aoms/1177704477}

\bibitem[{{Andreux} {et~al.}(2018){Andreux}, {Angles}, {Exarchakis}, {Leonarduzzi}, {Rochette}, {Thiry}, {Zarka}, {Mallat}, {And{\'e}n}, {Belilovsky}, {Bruna}, {Lostanlen}, {Hirn}, {Oyallon}, {Zhang}, {Cella}, \& {Eickenberg}}]{2018arXiv181211214A}
{Andreux}, M., {Angles}, T., {Exarchakis}, G., {et~al.} 2018, arXiv e-prints, arXiv:1812.11214.
\newblock \doarXiv{1812.11214}

\bibitem[{{Astropy Collaboration} {et~al.}(2013){Astropy Collaboration}, {Robitaille}, {Tollerud}, {Greenfield}, {Droettboom}, {Bray}, {Aldcroft}, {Davis}, {Ginsburg}, {Price-Whelan}, {Kerzendorf}, {Conley}, {Crighton}, {Barbary}, {Muna}, {Ferguson}, {Grollier}, {Parikh}, {Nair}, {Unther}, {Deil}, {Woillez}, {Conseil}, {Kramer}, {Turner}, {Singer}, {Fox}, {Weaver}, {Zabalza}, {Edwards}, {Azalee Bostroem}, {Burke}, {Casey}, {Crawford}, {Dencheva}, {Ely}, {Jenness}, {Labrie}, {Lim}, {Pierfederici}, {Pontzen}, {Ptak}, {Refsdal}, {Servillat}, \& {Streicher}}]{2013A&A...558A..33A}
{Astropy Collaboration}, {Robitaille}, T.~P., {Tollerud}, E.~J., {et~al.} 2013, \aap, 558, A33, \dodoi{10.1051/0004-6361/201322068}

\bibitem[{{Astropy Collaboration} {et~al.}(2018){Astropy Collaboration}, {Price-Whelan}, {Sip{\H{o}}cz}, {G{\"u}nther}, {Lim}, {Crawford}, {Conseil}, {Shupe}, {Craig}, {Dencheva}, {Ginsburg}, {VanderPlas}, {Bradley}, {P{\'e}rez-Su{\'a}rez}, {de Val-Borro}, {Aldcroft}, {Cruz}, {Robitaille}, {Tollerud}, {Ardelean}, {Babej}, {Bach}, {Bachetti}, {Bakanov}, {Bamford}, {Barentsen}, {Barmby}, {Baumbach}, {Berry}, {Biscani}, {Boquien}, {Bostroem}, {Bouma}, {Brammer}, {Bray}, {Breytenbach}, {Buddelmeijer}, {Burke}, {Calderone}, {Cano Rodr{\'\i}guez}, {Cara}, {Cardoso}, {Cheedella}, {Copin}, {Corrales}, {Crichton}, {D'Avella}, {Deil}, {Depagne}, {Dietrich}, {Donath}, {Droettboom}, {Earl}, {Erben}, {Fabbro}, {Ferreira}, {Finethy}, {Fox}, {Garrison}, {Gibbons}, {Goldstein}, {Gommers}, {Greco}, {Greenfield}, {Groener}, {Grollier}, {Hagen}, {Hirst}, {Homeier}, {Horton}, {Hosseinzadeh}, {Hu}, {Hunkeler}, {Ivezi{\'c}}, {Jain}, {Jenness}, {Kanarek}, {Kendrew}, {Kern}, {Kerzendorf}, {Khvalko}, {King}, {Kirkby}, {Kulkarni},
  {Kumar}, {Lee}, {Lenz}, {Littlefair}, {Ma}, {Macleod}, {Mastropietro}, {McCully}, {Montagnac}, {Morris}, {Mueller}, {Mumford}, {Muna}, {Murphy}, {Nelson}, {Nguyen}, {Ninan}, {N{\"o}the}, {Ogaz}, {Oh}, {Parejko}, {Parley}, {Pascual}, {Patil}, {Patil}, {Plunkett}, {Prochaska}, {Rastogi}, {Reddy Janga}, {Sabater}, {Sakurikar}, {Seifert}, {Sherbert}, {Sherwood-Taylor}, {Shih}, {Sick}, {Silbiger}, {Singanamalla}, {Singer}, {Sladen}, {Sooley}, {Sornarajah}, {Streicher}, {Teuben}, {Thomas}, {Tremblay}, {Turner}, {Terr{\'o}n}, {van Kerkwijk}, {de la Vega}, {Watkins}, {Weaver}, {Whitmore}, {Woillez}, {Zabalza}, \& {Astropy Contributors}}]{2018AJ....156..123A}
{Astropy Collaboration}, {Price-Whelan}, A.~M., {Sip{\H{o}}cz}, B.~M., {et~al.} 2018, \aj, 156, 123, \dodoi{10.3847/1538-3881/aabc4f}

\bibitem[{Bianco {et~al.}(2021)Bianco, Giri, Iliev, \& Mellema}]{10.1093/mnras/stab1518}
Bianco, M., Giri, S.~K., Iliev, I.~T., \& Mellema, G. 2021, Monthly Notices of the Royal Astronomical Society, 505, 3982, \dodoi{10.1093/mnras/stab1518}

\bibitem[{{Cheng} \& {M{\'e}nard}(2021)}]{2021arXiv211201288C}
{Cheng}, S., \& {M{\'e}nard}, B. 2021, arXiv e-prints, arXiv:2112.01288.
\newblock \doarXiv{2112.01288}

\bibitem[{{Cheng} {et~al.}(2020){Cheng}, {Ting}, {M{\'e}nard}, \& {Bruna}}]{cheng2020new}
{Cheng}, S., {Ting}, Y.-S., {M{\'e}nard}, B., \& {Bruna}, J. 2020, \mnras, 499, 5902, \dodoi{10.1093/mnras/staa3165}

\bibitem[{{Chung}(2022)}]{2022MNRAS.517.1625C}
{Chung}, D.~T. 2022, \mnras, 517, 1625, \dodoi{10.1093/mnras/stac2662}

\bibitem[{Cranmer {et~al.}(2020)Cranmer, Brehmer, \& Louppe}]{Cranmer30055}
Cranmer, K., Brehmer, J., \& Louppe, G. 2020, Proceedings of the National Academy of Sciences, 117, 30055, \dodoi{10.1073/pnas.1912789117}

\bibitem[{{de Oliveira-Costa} {et~al.}(2008){de Oliveira-Costa}, {Tegmark}, {Gaensler}, {Jonas}, {Landecker}, \& {Reich}}]{2008MNRAS.388..247D}
{de Oliveira-Costa}, A., {Tegmark}, M., {Gaensler}, B.~M., {et~al.} 2008, \mnras, 388, 247, \dodoi{10.1111/j.1365-2966.2008.13376.x}

\bibitem[{DeBoer {et~al.}(2017)DeBoer, Parsons, Aguirre, Alexander, Ali, Beardsley, Bernardi, Bowman, Bradley, Carilli, Cheng, de~Lera~Acedo, Dillon, Ewall-Wice, Fadana, Fagnoni, Fritz, Furlanetto, Glendenning, Greig, Grobbelaar, Hazelton, Hewitt, Hickish, Jacobs, Julius, Kariseb, Kohn, Lekalake, Liu, Loots, MacMahon, Malan, Malgas, Maree, Martinot, Mathison, Matsetela, Mesinger, Morales, Neben, Patra, Pieterse, Pober, Razavi-Ghods, Ringuette, Robnett, Rosie, Sell, Smith, Syce, Tegmark, Thyagarajan, Williams, \& Zheng}]{DeBoer2017}
DeBoer, D.~R., Parsons, A.~R., Aguirre, J.~E., {et~al.} 2017, Publications of the Astronomical Society of the Pacific, 129, 045001.
\newblock \url{http://stacks.iop.org/1538-3873/129/i=974/a=045001}

\bibitem[{{DeZoort} \& {Hanin}(2023)}]{2023arXiv230611668D}
{DeZoort}, G., \& {Hanin}, B. 2023, arXiv e-prints, arXiv:2306.11668, \dodoi{10.48550/arXiv.2306.11668}

\bibitem[{Eickenberg {et~al.}(2017)Eickenberg, Exarchakis, Hirn, \& Mallat}]{eickenberg2017solid}
Eickenberg, M., Exarchakis, G., Hirn, M., \& Mallat, S. 2017, in Proceedings of the 31st International Conference on Neural Information Processing Systems, NIPS'17 (Red Hook, NY, USA: Curran Associates Inc.), 6543–6552

\bibitem[{Eickenberg {et~al.}(2018)Eickenberg, Exarchakis, Hirn, Mallat, \& Thiry}]{eickenberg2018solid}
Eickenberg, M., Exarchakis, G., Hirn, M., Mallat, S., \& Thiry, L. 2018, The Journal of chemical physics, 148, 241732

\bibitem[{{Eickenberg} {et~al.}(2022){Eickenberg}, {Allys}, {Moradinezhad Dizgah}, {Lemos}, {Massara}, {Abidi}, {Hahn}, {Hassan}, {Regaldo-Saint Blancard}, {Ho}, {Mallat}, {And{\'e}n}, \& {Villaescusa-Navarro}}]{2022arXiv220407646E}
{Eickenberg}, M., {Allys}, E., {Moradinezhad Dizgah}, A., {et~al.} 2022, arXiv e-prints, arXiv:2204.07646, \dodoi{10.48550/arXiv.2204.07646}

\bibitem[{Furlanetto {et~al.}(2006)Furlanetto, Oh, \& Briggs}]{Furlanetto2006}
Furlanetto, S.~R., Oh, S.~P., \& Briggs, F.~H. 2006, Physics Reports, 433, 181, \dodoi{10.1016/j.physrep.2006.08.002}

\bibitem[{{Gauthier} {et~al.}(2021){Gauthier}, {Th{\'e}rien}, {Als{\`e}ne-Racicot}, {Chaudhary}, {Rish}, {Belilovsky}, {Eickenberg}, \& {Wolf}}]{2021arXiv210709539G}
{Gauthier}, S., {Th{\'e}rien}, B., {Als{\`e}ne-Racicot}, L., {et~al.} 2021, arXiv e-prints, arXiv:2107.09539, \dodoi{10.48550/arXiv.2107.09539}

\bibitem[{Germain {et~al.}(2015)Germain, Gregor, Murray, \& Larochelle}]{germain2015made}
Germain, M., Gregor, K., Murray, I., \& Larochelle, H. 2015, in International Conference on Machine Learning, 881--889

\bibitem[{{Gillet} {et~al.}(2019){Gillet}, {Mesinger}, {Greig}, {Liu}, \& {Ucci}}]{Gillet2019}
{Gillet}, N., {Mesinger}, A., {Greig}, B., {Liu}, A., \& {Ucci}, G. 2019, \mnras, 484, 282, \dodoi{10.1093/mnras/stz010}

\bibitem[{{Giri} \& {Mellema}(2021)}]{giri2020measuring}
{Giri}, S.~K., \& {Mellema}, G. 2021, \mnras, 505, 1863, \dodoi{10.1093/mnras/stab1320}

\bibitem[{{Giri} {et~al.}(2018){Giri}, {Mellema}, \& {Ghara}}]{2018MNRAS.479.5596G}
{Giri}, S.~K., {Mellema}, G., \& {Ghara}, R. 2018, \mnras, 479, 5596, \dodoi{10.1093/mnras/sty1786}

\bibitem[{Giri {et~al.}(2020)Giri, Mellema, \& Jensen}]{Giri2020}
Giri, S.~K., Mellema, G., \& Jensen, H. 2020, Journal of Open Source Software, 5, 2363, \dodoi{10.21105/joss.02363}

\bibitem[{Gneiting {et~al.}(2007)Gneiting, Balabdaoui, \& Raftery}]{gneiting2007probabilistic}
Gneiting, T., Balabdaoui, F., \& Raftery, A.~E. 2007, Journal of the Royal Statistical Society: Series B (Statistical Methodology), 69, 243, \dodoi{https://doi.org/10.1111/j.1467-9868.2007.00587.x}

\bibitem[{{G{\'o}rski} {et~al.}(2005){G{\'o}rski}, {Hivon}, {Banday}, {Wandelt}, {Hansen}, {Reinecke}, \& {Bartelmann}}]{2005ApJ...622..759G}
{G{\'o}rski}, K.~M., {Hivon}, E., {Banday}, A.~J., {et~al.} 2005, \apj, 622, 759, \dodoi{10.1086/427976}

\bibitem[{{Greig} \& {Mesinger}(2015)}]{2015MNRAS.449.4246G}
{Greig}, B., \& {Mesinger}, A. 2015, \mnras, 449, 4246, \dodoi{10.1093/mnras/stv571}

\bibitem[{{Greig} \& {Mesinger}(2017)}]{2017MNRAS.472.2651G}
---. 2017, \mnras, 472, 2651, \dodoi{10.1093/mnras/stx2118}

\bibitem[{{Greig} \& {Mesinger}(2018)}]{Greig2018}
---. 2018, \mnras, 477, 3217, \dodoi{10.1093/mnras/sty796}

\bibitem[{{Greig} {et~al.}(2017){Greig}, {Mesinger}, {Haiman}, \& {Simcoe}}]{2017MNRAS.466.4239G}
{Greig}, B., {Mesinger}, A., {Haiman}, Z., \& {Simcoe}, R.~A. 2017, \mnras, 466, 4239, \dodoi{10.1093/mnras/stw3351}

\bibitem[{{Greig} {et~al.}(2022){Greig}, {Ting}, \& {Kaurov}}]{2022arXiv220402544G}
{Greig}, B., {Ting}, Y.-S., \& {Kaurov}, A.~A. 2022, arXiv e-prints, arXiv:2204.02544.
\newblock \doarXiv{2204.02544}

\bibitem[{Harris {et~al.}(2020)Harris, Millman, van~der Walt, Gommers, Virtanen, Cournapeau, Wieser, Taylor, Berg, Smith, Kern, Picus, Hoyer, van Kerkwijk, Brett, Haldane, del R{\'{i}}o, Wiebe, Peterson, G{\'{e}}rard-Marchant, Sheppard, Reddy, Weckesser, Abbasi, Gohlke, \& Oliphant}]{harris2020array}
Harris, C.~R., Millman, K.~J., van~der Walt, S.~J., {et~al.} 2020, Nature, 585, 357, \dodoi{10.1038/s41586-020-2649-2}

\bibitem[{{Harrison} {et~al.}(2015){Harrison}, {Sutton}, {Carvalho}, \& {Hobson}}]{harrison2015validation}
{Harrison}, D., {Sutton}, D., {Carvalho}, P., \& {Hobson}, M. 2015, \mnras, 451, 2610, \dodoi{10.1093/mnras/stv1110}

\bibitem[{Hunter(2007)}]{Hunter:2007}
Hunter, J.~D. 2007, Computing in Science \& Engineering, 9, 90, \dodoi{10.1109/MCSE.2007.55}

\bibitem[{{Intema} {et~al.}(2017){Intema}, {Jagannathan}, {Mooley}, \& {Frail}}]{2017A&A...598A..78I}
{Intema}, H.~T., {Jagannathan}, P., {Mooley}, K.~P., \& {Frail}, D.~A. 2017, \aap, 598, A78, \dodoi{10.1051/0004-6361/201628536}

\bibitem[{{Jacobs} {et~al.}(2015){Jacobs}, {Pober}, {Parsons}, {Aguirre}, {Ali}, {Bowman}, {Bradley}, {Carilli}, {DeBoer}, {Dexter}, {Gugliucci}, {Klima}, {Liu}, {MacMahon}, {Manley}, {Moore}, {Stefan}, \& {Walbrugh}}]{2015ApJ...801...51J}
{Jacobs}, D.~C., {Pober}, J.~C., {Parsons}, A.~R., {et~al.} 2015, \apj, 801, 51, \dodoi{10.1088/0004-637X/801/1/51}

\bibitem[{{Jeffrey} {et~al.}(2022){Jeffrey}, {Boulanger}, {Wandelt}, {Regaldo-Saint Blancard}, {Allys}, \& {Levrier}}]{2022MNRAS.510L...1J}
{Jeffrey}, N., {Boulanger}, F., {Wandelt}, B.~D., {et~al.} 2022, \mnras, 510, L1, \dodoi{10.1093/mnrasl/slab120}

\bibitem[{{Kauderer-Abrams}(2017)}]{2018arXiv180101450K}
{Kauderer-Abrams}, E. 2017, arXiv e-prints, arXiv:1801.01450.
\newblock \doarXiv{1801.01450}

\bibitem[{Kolmogorov(1992)}]{Kolmogorov1992}
Kolmogorov, A. 1992, On the Empirical Determination of a Distribution Function, ed. S.~Kotz \& N.~L. Johnson (New York, NY: Springer New York), 106--113, \dodoi{10.1007/978-1-4612-4380-9_10}

\bibitem[{{Legin} {et~al.}(2023){Legin}, {Ho}, {Lemos}, {Perreault-Levasseur}, {Ho}, {Hezaveh}, \& {Wandelt}}]{2023arXiv230403788L}
{Legin}, R., {Ho}, M., {Lemos}, P., {et~al.} 2023, arXiv e-prints, arXiv:2304.03788, \dodoi{10.48550/arXiv.2304.03788}

\bibitem[{Lewis(2019)}]{Lewis:2019xzd}
Lewis, A. 2019.
\newblock \doarXiv{1910.13970}

\bibitem[{{Luo}(2022)}]{2022arXiv220811970L}
{Luo}, C. 2022, arXiv e-prints, arXiv:2208.11970, \dodoi{10.48550/arXiv.2208.11970}

\bibitem[{Mallat(2012{\natexlab{a}})}]{mallat2012group}
Mallat, S. 2012{\natexlab{a}}, Communications on Pure and Applied Mathematics, 65, 1331, \dodoi{https://doi.org/10.1002/cpa.21413}

\bibitem[{Mallat(2012{\natexlab{b}})}]{https://doi.org/10.1002/cpa.21413}
---. 2012{\natexlab{b}}, Communications on Pure and Applied Mathematics, 65, 1331, \dodoi{https://doi.org/10.1002/cpa.21413}

\bibitem[{Mallat {et~al.}(2019)Mallat, Zhang, \& Rochette}]{10.1093/imaiai/iaz019}
Mallat, S., Zhang, S., \& Rochette, G. 2019, Information and Inference: A Journal of the IMA, 9, 721, \dodoi{10.1093/imaiai/iaz019}

\bibitem[{{Mao} {et~al.}(2012){Mao}, {Shapiro}, {Mellema}, {Iliev}, {Koda}, \& {Ahn}}]{2012MNRAS.422..926M}
{Mao}, Y., {Shapiro}, P.~R., {Mellema}, G., {et~al.} 2012, \mnras, 422, 926, \dodoi{10.1111/j.1365-2966.2012.20471.x}

\bibitem[{{Masui} {et~al.}(2013){Masui}, {Switzer}, {Banavar}, {Bandura}, {Blake}, {Calin}, {Chang}, {Chen}, {Li}, {Liao}, {Natarajan}, {Pen}, {Peterson}, {Shaw}, \& {Voytek}}]{masui2013measurement}
{Masui}, K.~W., {Switzer}, E.~R., {Banavar}, N., {et~al.} 2013, \apjl, 763, L20, \dodoi{10.1088/2041-8205/763/1/L20}

\bibitem[{{Mellema} {et~al.}(2013){Mellema}, {Koopmans}, {Abdalla}, {Bernardi}, {Ciardi}, {Daiboo}, {de Bruyn}, {Datta}, {Falcke}, {Ferrara}, {Iliev}, {Iocco}, {Jeli{\'c}}, {Jensen}, {Joseph}, {Labroupoulos}, {Meiksin}, {Mesinger}, {Offringa}, {Pandey}, {Pritchard}, {Santos}, {Schwarz}, {Semelin}, {Vedantham}, {Yatawatta}, \& {Zaroubi}}]{Mellema2013}
{Mellema}, G., {Koopmans}, L. V.~E., {Abdalla}, F.~A., {et~al.} 2013, Experimental Astronomy, 36, 235, \dodoi{10.1007/s10686-013-9334-5}

\bibitem[{{Mertens} {et~al.}(2020){Mertens}, {Mevius}, {Koopmans}, {Offringa}, {Mellema}, {Zaroubi}, {Brentjens}, {Gan}, {Gehlot}, {Pandey}, {Sardarabadi}, {Vedantham}, {Yatawatta}, {Asad}, {Ciardi}, {Chapman}, {Gazagnes}, {Ghara}, {Ghosh}, {Giri}, {Iliev}, {Jeli{\'c}}, {Kooistra}, {Mondal}, {Schaye}, \& {Silva}}]{2020MNRAS.493.1662M}
{Mertens}, F.~G., {Mevius}, M., {Koopmans}, L.~V.~E., {et~al.} 2020, \mnras, 493, 1662, \dodoi{10.1093/mnras/staa327}

\bibitem[{{Mesinger} \& {Furlanetto}(2007)}]{Mesinger2007}
{Mesinger}, A., \& {Furlanetto}, S. 2007, \apj, 669, 663, \dodoi{10.1086/521806}

\bibitem[{Mesinger {et~al.}(2011)Mesinger, Furlanetto, \& Cen}]{Mesinger2011}
Mesinger, A., Furlanetto, S., \& Cen, R. 2011, \mnras, 411, 955, \dodoi{10.1111/j.1365-2966.2010.17731.x}

\bibitem[{{Mucesh} {et~al.}(2021){Mucesh}, {Hartley}, {Palmese}, {Lahav}, {Whiteway}, {Bluck}, {Alarcon}, {Amon}, {Bechtol}, {Bernstein}, {Carnero Rosell}, {Carrasco Kind}, {Choi}, {Eckert}, {Everett}, {Gruen}, {Gruendl}, {Harrison}, {Huff}, {Kuropatkin}, {Sevilla-Noarbe}, {Sheldon}, {Yanny}, {Aguena}, {Allam}, {Bacon}, {Bertin}, {Bhargava}, {Brooks}, {Carretero}, {Castander}, {Conselice}, {Costanzi}, {Crocce}, {da Costa}, {Pereira}, {De Vicente}, {Desai}, {Diehl}, {Drlica-Wagner}, {Evrard}, {Ferrero}, {Flaugher}, {Fosalba}, {Frieman}, {Garc{\'\i}a-Bellido}, {Gaztanaga}, {Gerdes}, {Gschwend}, {Gutierrez}, {Hinton}, {Hollowood}, {Honscheid}, {James}, {Kuehn}, {Lima}, {Lin}, {Maia}, {Melchior}, {Menanteau}, {Miquel}, {Morgan}, {Paz-Chinch{\'o}n}, {Plazas}, {Sanchez}, {Scarpine}, {Schubnell}, {Serrano}, {Smith}, {Suchyta}, {Tarle}, {Thomas}, {To}, {Varga}, {Wilkinson}, \& {DES Collaboration}}]{mucesh2021machine}
{Mucesh}, S., {Hartley}, W.~G., {Palmese}, A., {et~al.} 2021, \mnras, 502, 2770, \dodoi{10.1093/mnras/stab164}

\bibitem[{{Neutsch} {et~al.}(2022){Neutsch}, {Heneka}, \& {Br{\"u}ggen}}]{2022MNRAS.511.3446N}
{Neutsch}, S., {Heneka}, C., \& {Br{\"u}ggen}, M. 2022, \mnras, 511, 3446, \dodoi{10.1093/mnras/stac218}

\bibitem[{{Paciga} {et~al.}(2013){Paciga}, {Albert}, {Bandura}, {Chang}, {Gupta}, {Hirata}, {Odegova}, {Pen}, {Peterson}, {Roy}, {Shaw}, {Sigurdson}, \& {Voytek}}]{2013MNRAS.433..639P}
{Paciga}, G., {Albert}, J.~G., {Bandura}, K., {et~al.} 2013, \mnras, 433, 639, \dodoi{10.1093/mnras/stt753}

\bibitem[{{Papamakarios}(2019)}]{2019arXiv191013233P}
{Papamakarios}, G. 2019, arXiv e-prints, arXiv:1910.13233.
\newblock \doarXiv{1910.13233}

\bibitem[{Papamakarios {et~al.}(2021)Papamakarios, Nalisnick, Rezende, Mohamed, \& Lakshminarayanan}]{JMLR:v22:19-1028}
Papamakarios, G., Nalisnick, E., Rezende, D.~J., Mohamed, S., \& Lakshminarayanan, B. 2021, Journal of Machine Learning Research, 22, 1.
\newblock \url{http://jmlr.org/papers/v22/19-1028.html}

\bibitem[{Papamakarios {et~al.}(2017)Papamakarios, Pavlakou, \& Murray}]{papamakarios2017masked}
Papamakarios, G., Pavlakou, T., \& Murray, I. 2017, in Proceedings of the 31st International Conference on Neural Information Processing Systems, NIPS'17 (Red Hook, NY, USA: Curran Associates Inc.), 2335

\bibitem[{Parsons {et~al.}(2010)Parsons, Backer, Foster, Wright, Bradley, Gugliucci, Parashare, Benoit, Aguirre, Jacobs, Carilli, Herne, Lynch, Manley, \& Werthimer}]{Parsons2010}
Parsons, A.~R., Backer, D.~C., Foster, G.~S., {et~al.} 2010, \apj, 139, 1468.
\newblock \url{http://stacks.iop.org/1538-3881/139/i=4/a=1468}

\bibitem[{Parsons {et~al.}(2014)Parsons, Liu, Aguirre, Ali, Bradley, Carilli, DeBoer, Dexter, Gugliucci, Jacobs, Klima, MacMahon, Manley, Moore, Pober, Stefan, \& Walbrugh}]{Parsons_2014}
Parsons, A.~R., Liu, A., Aguirre, J.~E., {et~al.} 2014, The Astrophysical Journal, 788, 106, \dodoi{10.1088/0004-637X/788/2/106}

\bibitem[{{Pedersen} {et~al.}(2022){Pedersen}, {Ho}, \& {Eickenberg}}]{2022mla..confE..40P}
{Pedersen}, C., {Ho}, S., \& {Eickenberg}, M. 2022, in Machine Learning for Astrophysics, 40

\bibitem[{Pedregosa {et~al.}(2011)Pedregosa, Varoquaux, Gramfort, Michel, Thirion, Grisel, Blondel, Prettenhofer, Weiss, Dubourg, Vanderplas, Passos, Cournapeau, Brucher, Perrot, \& Duchesnay}]{Scikit-learn}
Pedregosa, F., Varoquaux, G., Gramfort, A., {et~al.} 2011, Journal of Machine Learning Research, 12, 2825

\bibitem[{{Planck Collaboration} {et~al.}(2016){Planck Collaboration}, {Ade}, {Aghanim}, {Arnaud}, {Ashdown}, {Aumont}, {Baccigalupi}, {Banday}, {Barreiro}, {Bartlett}, {Bartolo}, {Battaner}, {Battye}, {Benabed}, {Beno{\^\i}t}, {Benoit-L{\'e}vy}, {Bernard}, {Bersanelli}, {Bielewicz}, {Bock}, {Bonaldi}, {Bonavera}, {Bond}, {Borrill}, {Bouchet}, {Boulanger}, {Bucher}, {Burigana}, {Butler}, {Calabrese}, {Cardoso}, {Catalano}, {Challinor}, {Chamballu}, {Chary}, {Chiang}, {Chluba}, {Christensen}, {Church}, {Clements}, {Colombi}, {Colombo}, {Combet}, {Coulais}, {Crill}, {Curto}, {Cuttaia}, {Danese}, {Davies}, {Davis}, {de Bernardis}, {de Rosa}, {de Zotti}, {Delabrouille}, {D{\'e}sert}, {Di Valentino}, {Dickinson}, {Diego}, {Dolag}, {Dole}, {Donzelli}, {Dor{\'e}}, {Douspis}, {Ducout}, {Dunkley}, {Dupac}, {Efstathiou}, {Elsner}, {En{\ss}lin}, {Eriksen}, {Farhang}, {Fergusson}, {Finelli}, {Forni}, {Frailis}, {Fraisse}, {Franceschi}, {Frejsel}, {Galeotta}, {Galli}, {Ganga}, {Gauthier}, {Gerbino}, {Ghosh}, {Giard},
  {Giraud-H{\'e}raud}, {Giusarma}, {Gjerl{\o}w}, {Gonz{\'a}lez-Nuevo}, {G{\'o}rski}, {Gratton}, {Gregorio}, {Gruppuso}, {Gudmundsson}, {Hamann}, {Hansen}, {Hanson}, {Harrison}, {Helou}, {Henrot-Versill{\'e}}, {Hern{\'a}ndez-Monteagudo}, {Herranz}, {Hildebrandt}, {Hivon}, {Hobson}, {Holmes}, {Hornstrup}, {Hovest}, {Huang}, {Huffenberger}, {Hurier}, {Jaffe}, {Jaffe}, {Jones}, {Juvela}, {Keih{\"a}nen}, {Keskitalo}, {Kisner}, {Kneissl}, {Knoche}, {Knox}, {Kunz}, {Kurki-Suonio}, {Lagache}, {L{\"a}hteenm{\"a}ki}, {Lamarre}, {Lasenby}, {Lattanzi}, {Lawrence}, {Leahy}, {Leonardi}, {Lesgourgues}, {Levrier}, {Lewis}, {Liguori}, {Lilje}, {Linden-V{\o}rnle}, {L{\'o}pez-Caniego}, {Lubin}, {Mac{\'\i}as-P{\'e}rez}, {Maggio}, {Maino}, {Mandolesi}, {Mangilli}, {Marchini}, {Maris}, {Martin}, {Martinelli}, {Mart{\'\i}nez-Gonz{\'a}lez}, {Masi}, {Matarrese}, {McGehee}, {Meinhold}, {Melchiorri}, {Melin}, {Mendes}, {Mennella}, {Migliaccio}, {Millea}, {Mitra}, {Miville-Desch{\^e}nes}, {Moneti}, {Montier}, {Morgante}, {Mortlock},
  {Moss}, {Munshi}, {Murphy}, {Naselsky}, {Nati}, {Natoli}, {Netterfield}, {N{\o}rgaard-Nielsen}, {Noviello}, {Novikov}, {Novikov}, {Oxborrow}, {Paci}, {Pagano}, {Pajot}, {Paladini}, {Paoletti}, {Partridge}, {Pasian}, {Patanchon}, {Pearson}, {Perdereau}, {Perotto}, {Perrotta}, {Pettorino}, {Piacentini}, {Piat}, {Pierpaoli}, {Pietrobon}, {Plaszczynski}, {Pointecouteau}, {Polenta}, {Popa}, {Pratt}, {Pr{\'e}zeau}, {Prunet}, {Puget}, {Rachen}, {Reach}, {Rebolo}, {Reinecke}, {Remazeilles}, {Renault}, {Renzi}, {Ristorcelli}, {Rocha}, {Rosset}, {Rossetti}, {Roudier}, {Rouill{\'e} d'Orfeuil}, {Rowan-Robinson}, {Rubi{\~n}o-Mart{\'\i}n}, {Rusholme}, {Said}, {Salvatelli}, {Salvati}, {Sandri}, {Santos}, {Savelainen}, {Savini}, {Scott}, {Seiffert}, {Serra}, {Shellard}, {Spencer}, {Spinelli}, {Stolyarov}, {Stompor}, {Sudiwala}, {Sunyaev}, {Sutton}, {Suur-Uski}, {Sygnet}, {Tauber}, {Terenzi}, {Toffolatti}, {Tomasi}, {Tristram}, {Trombetti}, {Tucci}, {Tuovinen}, {T{\"u}rler}, {Umana}, {Valenziano}, {Valiviita}, {Van Tent},
  {Vielva}, {Villa}, {Wade}, {Wandelt}, {Wehus}, {White}, {White}, {Wilkinson}, {Yvon}, {Zacchei}, \& {Zonca}}]{ade2016planck}
{Planck Collaboration}, {Ade}, P.~A.~R., {Aghanim}, N., {et~al.} 2016, \aap, 594, A13, \dodoi{10.1051/0004-6361/201525830}

\bibitem[{{Prelogovi{\'c}} \& {Mesinger}(2023)}]{2023arXiv230503074P}
{Prelogovi{\'c}}, D., \& {Mesinger}, A. 2023, arXiv e-prints, arXiv:2305.03074, \dodoi{10.48550/arXiv.2305.03074}

\bibitem[{{Prelogovi{\'c}} {et~al.}(2022){Prelogovi{\'c}}, {Mesinger}, {Murray}, {Fiameni}, \& {Gillet}}]{2022MNRAS.509.3852P}
{Prelogovi{\'c}}, D., {Mesinger}, A., {Murray}, S., {Fiameni}, G., \& {Gillet}, N. 2022, \mnras, 509, 3852, \dodoi{10.1093/mnras/stab3215}

\bibitem[{Ramachandran \& Varoquaux(2011)}]{ramachandran2011mayavi}
Ramachandran, P., \& Varoquaux, G. 2011, Computing in Science and Engg., 13, 40–51, \dodoi{10.1109/MCSE.2011.35}

\bibitem[{{Saydjari} {et~al.}(2021){Saydjari}, {Portillo}, {Slepian}, {Kahraman}, {Burkhart}, \& {Finkbeiner}}]{2021ApJ...910..122S}
{Saydjari}, A.~K., {Portillo}, S. K.~N., {Slepian}, Z., {et~al.} 2021, \apj, 910, 122, \dodoi{10.3847/1538-4357/abe46d}

\bibitem[{Semih~Kayhan \& van Gemert(2020)}]{9156444}
Semih~Kayhan, O., \& van Gemert, J.~C. 2020, in 2020 IEEE/CVF Conference on Computer Vision and Pattern Recognition (CVPR) (Piscataway, NJ: IEEE), 14262, \dodoi{10.1109/CVPR42600.2020.01428}

\bibitem[{Stewart(1993)}]{stewart1993early}
Stewart, G.~W. 1993, SIAM Rev., 35, 551–566, \dodoi{10.1137/1035134}

\bibitem[{{Sui} {et~al.}(2023){Sui}, {Zhao}, {Jing}, \& {Mao}}]{2023arXiv230704994S}
{Sui}, C., {Zhao}, X., {Jing}, T., \& {Mao}, Y. 2023, arXiv e-prints, arXiv:2307.04994, \dodoi{10.48550/arXiv.2307.04994}

\bibitem[{Tejero-Cantero {et~al.}(2020)Tejero-Cantero, Boelts, Deistler, Lueckmann, Durkan, Gonçalves, Greenberg, \& Macke}]{tejero-cantero2020sbi}
Tejero-Cantero, A., Boelts, J., Deistler, M., {et~al.} 2020, Journal of Open Source Software, 5, 2505, \dodoi{10.21105/joss.02505}

\bibitem[{Tingay {et~al.}(2013)Tingay, Goeke, Bowman, Emrich, Ord, Mitchell, Morales, Booler, Crosse, Wayth, \& et~al.}]{Tingay2013}
Tingay, S.~J., Goeke, R., Bowman, J.~D., {et~al.} 2013, Publications of the Astronomical Society of Australia, 30, e007, \dodoi{10.1017/pasa.2012.007}

\bibitem[{Trott(2016)}]{10.1093/mnras/stw1310}
Trott, C.~M. 2016, \mnras, 461, 126, \dodoi{10.1093/mnras/stw1310}

\bibitem[{{Trott} {et~al.}(2020){Trott}, {Jordan}, {Midgley}, {Barry}, {Greig}, {Pindor}, {Cook}, {Sleap}, {Tingay}, {Ung}, {Hancock}, {Williams}, {Bowman}, {Byrne}, {Chokshi}, {Hazelton}, {Hasegawa}, {Jacobs}, {Joseph}, {Li}, {Line}, {Lynch}, {McKinley}, {Mitchell}, {Morales}, {Ouchi}, {Pober}, {Rahimi}, {Takahashi}, {Wayth}, {Webster}, {Wilensky}, {Wyithe}, {Yoshiura}, {Zhang}, \& {Zheng}}]{2020MNRAS.493.4711T}
{Trott}, C.~M., {Jordan}, C.~H., {Midgley}, S., {et~al.} 2020, \mnras, 493, 4711, \dodoi{10.1093/mnras/staa414}

\bibitem[{{Valogiannis} \& {Dvorkin}(2022)}]{2022PhRvD.105j3534V}
{Valogiannis}, G., \& {Dvorkin}, C. 2022, \prd, 105, 103534, \dodoi{10.1103/PhysRevD.105.103534}

\bibitem[{{van Haarlem} {et~al.}(2013){van Haarlem}, {Wise}, {Gunst}, {Heald}, {McKean}, {Hessels}, {de Bruyn}, {Nijboer}, {Swinbank}, {Fallows}, {Brentjens}, {Nelles}, {Beck}, {Falcke}, {Fender}, {H{\"o}randel}, {Koopmans}, {Mann}, {Miley}, {R{\"o}ttgering}, {Stappers}, {Wijers}, {Zaroubi}, {van den Akker}, {Alexov}, {Anderson}, {Anderson}, {van Ardenne}, {Arts}, {Asgekar}, {Avruch}, {Batejat}, {B{\"a}hren}, {Bell}, {Bell}, {van Bemmel}, {Bennema}, {Bentum}, {Bernardi}, {Best}, {B{\^\i}rzan}, {Bonafede}, {Boonstra}, {Braun}, {Bregman}, {Breitling}, {van de Brink}, {Broderick}, {Broekema}, {Brouw}, {Br{\"u}ggen}, {Butcher}, {van Cappellen}, {Ciardi}, {Coenen}, {Conway}, {Coolen}, {Corstanje}, {Damstra}, {Davies}, {Deller}, {Dettmar}, {van Diepen}, {Dijkstra}, {Donker}, {Doorduin}, {Dromer}, {Drost}, {van Duin}, {Eisl{\"o}ffel}, {van Enst}, {Ferrari}, {Frieswijk}, {Gankema}, {Garrett}, {de Gasperin}, {Gerbers}, {de Geus}, {Grie{\ss}meier}, {Grit}, {Gruppen}, {Hamaker}, {Hassall}, {Hoeft}, {Holties},
  {Horneffer}, {van der Horst}, {van Houwelingen}, {Huijgen}, {Iacobelli}, {Intema}, {Jackson}, {Jelic}, {de Jong}, {Juette}, {Kant}, {Karastergiou}, {Koers}, {Kollen}, {Kondratiev}, {Kooistra}, {Koopman}, {Koster}, {Kuniyoshi}, {Kramer}, {Kuper}, {Lambropoulos}, {Law}, {van Leeuwen}, {Lemaitre}, {Loose}, {Maat}, {Macario}, {Markoff}, {Masters}, {McFadden}, {McKay-Bukowski}, {Meijering}, {Meulman}, {Mevius}, {Middelberg}, {Millenaar}, {Miller-Jones}, {Mohan}, {Mol}, {Morawietz}, {Morganti}, {Mulcahy}, {Mulder}, {Munk}, {Nieuwenhuis}, {van Nieuwpoort}, {Noordam}, {Norden}, {Noutsos}, {Offringa}, {Olofsson}, {Omar}, {Orr{\'u}}, {Overeem}, {Paas}, {Pandey-Pommier}, {Pandey}, {Pizzo}, {Polatidis}, {Rafferty}, {Rawlings}, {Reich}, {de Reijer}, {Reitsma}, {Renting}, {Riemers}, {Rol}, {Romein}, {Roosjen}, {Ruiter}, {Scaife}, {van der Schaaf}, {Scheers}, {Schellart}, {Schoenmakers}, {Schoonderbeek}, {Serylak}, {Shulevski}, {Sluman}, {Smirnov}, {Sobey}, {Spreeuw}, {Steinmetz}, {Sterks}, {Stiepel}, {Stuurwold},
  {Tagger}, {Tang}, {Tasse}, {Thomas}, {Thoudam}, {Toribio}, {van der Tol}, {Usov}, {van Veelen}, {van der Veen}, {ter Veen}, {Verbiest}, {Vermeulen}, {Vermaas}, {Vocks}, {Vogt}, {de Vos}, {van der Wal}, {van Weeren}, {Weggemans}, {Weltevrede}, {White}, {Wijnholds}, {Wilhelmsson}, {Wucknitz}, {Yatawatta}, {Zarka}, {Zensus}, \& {van Zwieten}}]{Haarlem2013}
{van Haarlem}, M.~P., {Wise}, M.~W., {Gunst}, A.~W., {et~al.} 2013, \aap, 556, A2, \dodoi{10.1051/0004-6361/201220873}

\bibitem[{Van~Rossum \& Drake(2009)}]{py3}
Van~Rossum, G., \& Drake, F.~L. 2009, Python 3 Reference Manual (Scotts Valley, CA: CreateSpace)

\bibitem[{Van~Rossum \& Drake~Jr(1995)}]{van1995python}
Van~Rossum, G., \& Drake~Jr, F.~L. 1995, Python reference manual (Centrum voor Wiskunde en Informatica Amsterdam)

\bibitem[{Virtanen {et~al.}(2020)Virtanen, Gommers, Oliphant, Haberland, Reddy, Cournapeau, Burovski, Peterson, Weckesser, Bright, {van der Walt}, Brett, Wilson, Millman, Mayorov, Nelson, Jones, Kern, Larson, Carey, Polat, Feng, Moore, {VanderPlas}, Laxalde, Perktold, Cimrman, Henriksen, Quintero, Harris, Archibald, Ribeiro, Pedregosa, {van Mulbregt}, \& {SciPy 1.0 Contributors}}]{2020SciPy-NMeth}
Virtanen, P., Gommers, R., Oliphant, T.~E., {et~al.} 2020, Nature Methods, 17, 261, \dodoi{10.1038/s41592-019-0686-2}

\bibitem[{Waskom(2021)}]{Waskom2021}
Waskom, M.~L. 2021, Journal of Open Source Software, 6, 3021, \dodoi{10.21105/joss.03021}

\bibitem[{Weiler {et~al.}(2018{\natexlab{a}})Weiler, Geiger, Welling, Boomsma, \& Cohen}]{NEURIPS2018_488e4104}
Weiler, M., Geiger, M., Welling, M., Boomsma, W., \& Cohen, T.~S. 2018{\natexlab{a}}, in Advances in Neural Information Processing Systems, ed. S.~Bengio, H.~Wallach, H.~Larochelle, K.~Grauman, N.~Cesa-Bianchi, \& R.~Garnett, Vol.~31 (Curran Associates, Inc.).
\newblock \url{https://proceedings.neurips.cc/paper_files/paper/2018/file/488e4104520c6aab692863cc1dba45af-Paper.pdf}

\bibitem[{Weiler {et~al.}(2018{\natexlab{b}})Weiler, Hamprecht, \& Storath}]{8578193}
Weiler, M., Hamprecht, F.~A., \& Storath, M. 2018{\natexlab{b}}, in 2018 IEEE/CVF Conference on Computer Vision and Pattern Recognition, 849--858, \dodoi{10.1109/CVPR.2018.00095}

\bibitem[{Wolz {et~al.}(2015)Wolz, Abdalla, Alonso, Blake, Bull, Chang, Ferreira, Kuo, Santos, \& Shaw}]{wolz2015foreground}
Wolz, L., Abdalla, F.~B., Alonso, D., {et~al.} 2015, PoS, AASKA14, 035, \dodoi{10.22323/1.215.0035}

\bibitem[{{Yoshiura} {et~al.}(2021){Yoshiura}, {Pindor}, {Line}, {Barry}, {Trott}, {Beardsley}, {Bowman}, {Byrne}, {Chokshi}, {Hazelton}, {Hasegawa}, {Howard}, {Greig}, {Jacobs}, {Jordan}, {Joseph}, {Kolopanis}, {Lynch}, {McKinley}, {Mitchell}, {Morales}, {Murray}, {Pober}, {Rahimi}, {Takahashi}, {Tingay}, {Wayth}, {Webster}, {Wilensky}, {Wyithe}, {Zhang}, \& {Zheng}}]{2021MNRAS.505.4775Y}
{Yoshiura}, S., {Pindor}, B., {Line}, J.~L.~B., {et~al.} 2021, \mnras, 505, 4775, \dodoi{10.1093/mnras/stab1560}

\bibitem[{{Zhao} {et~al.}(2022{\natexlab{a}}){Zhao}, {Mao}, {Cheng}, \& {Wandelt}}]{2021arXiv210503344Z}
{Zhao}, X., {Mao}, Y., {Cheng}, C., \& {Wandelt}, B.~D. 2022{\natexlab{a}}, \apj, 926, 151, \dodoi{10.3847/1538-4357/ac457d}

\bibitem[{{Zhao} {et~al.}(2022{\natexlab{b}}){Zhao}, {Mao}, \& {Wandelt}}]{zhao21cmdelfi}
{Zhao}, X., {Mao}, Y., \& {Wandelt}, B.~D. 2022{\natexlab{b}}, arXiv e-prints, arXiv:2203.15734.
\newblock \doarXiv{2203.15734}

\bibitem[{{Zhao} {et~al.}(2023{\natexlab{a}}){Zhao}, {Ting}, {Diao}, \& {Mao}}]{2023MNRAS.526.1699Z}
{Zhao}, X., {Ting}, Y.-S., {Diao}, K., \& {Mao}, Y. 2023{\natexlab{a}}, \mnras, 526, 1699, \dodoi{10.1093/mnras/stad2778}

\bibitem[{{Zhao} {et~al.}(2023{\natexlab{b}}){Zhao}, {Zuo}, \& {Mao}}]{2023arXiv230709530Z}
{Zhao}, X., {Zuo}, S., \& {Mao}, Y. 2023{\natexlab{b}}, arXiv e-prints, arXiv:2307.09530, \dodoi{10.48550/arXiv.2307.09530}

\bibitem[{{Zheng} {et~al.}(2017){Zheng}, {Tegmark}, {Dillon}, {Kim}, {Liu}, {Neben}, {Jonas}, {Reich}, \& {Reich}}]{zheng2017improved}
{Zheng}, H., {Tegmark}, M., {Dillon}, J.~S., {et~al.} 2017, \mnras, 464, 3486, \dodoi{10.1093/mnras/stw2525}

\bibitem[{Ziegel \& Gneiting(2014)}]{10.1214/14-EJS964}
Ziegel, J.~F., \& Gneiting, T. 2014, Electronic Journal of Statistics, 8, 2619 , \dodoi{10.1214/14-EJS964}

\bibitem[{Zonca {et~al.}(2019)Zonca, Singer, Lenz, Reinecke, Rosset, Hivon, \& Gorski}]{Zonca2019}
Zonca, A., Singer, L., Lenz, D., {et~al.} 2019, Journal of Open Source Software, 4, 1298, \dodoi{10.21105/joss.01298}

\bibitem[{Željko Ivezić {et~al.}(2014)Željko Ivezić, Connolly, VanderPlas, \& Gray}]{Ivezic2014}
Željko Ivezić, Connolly, A.~J., VanderPlas, J.~T., \& Gray, A. 2014, Statistics, Data Mining, and Machine Learning in Astronomy: A Practical Python Guide for the Analysis of Survey Data (Princeton University Press), \dodoi{doi:10.1515/9781400848911}

\end{thebibliography}

\end{document}